\pdfoutput=1

\documentclass[a4paper,fleqn]{cas-sc}

\usepackage[numbers]{natbib}

\usepackage{siunitx}
\usepackage{algorithm}
\usepackage{algpseudocode}
\usepackage{subcaption}
\usepackage{enumerate}
\usepackage{placeins}
\usepackage{verbatim}
\usepackage{threeparttable}
\usepackage{float}
\usepackage{tabularx}
\usepackage{pifont}        %
\usepackage{wasysym}       %
\usepackage[table]{xcolor}

\newcommand{\yes}{\textcolor{green!60!black}{\ding{51}}}        %
\renewcommand{\part}{\textcolor{orange!90!black}{\LEFTcircle}}  %
\newcommand{\no}{\textcolor{gray!70}{\ding{55}}}                %

\newcommand{\eulogo}[1]{%
  \IfFileExists{#1.pdf}{\includegraphics[height=1.7cm]{#1.pdf}}{%
   \IfFileExists{#1.png}{\includegraphics[height=1.7cm]{#1.png}}{%
    \IfFileExists{#1.jpg}{\includegraphics[height=1.7cm]{#1.jpg}}{%
     \fbox{\parbox[c][1.6cm][c]{5cm}{\centering\footnotesize EU / EuroHPC logo\\(image file to be added)}}}}}%
}

\ExplSyntaxOn
\RenewDocumentCommand \printorcid { }
  {
    \seq_if_empty:NF \g_stm_orcid_seq
      {
        \group_begin:
          \tex_let:D \thefootnote \relax
          \footnotetext
            {
              \raggedright
              \textsc{orcid}(s):\c_space_token
              \seq_use:Nn \g_stm_orcid_seq { ;~ }
            }
        \group_end:
      }
  }
\ExplSyntaxOff

\ExplSyntaxOn
\cs_set:Npn \__first_footerline: { }
\ExplSyntaxOff

\begin{document}
\let\WriteBookmarks\relax

\renewcommand{\topfraction}{0.92}
\renewcommand{\bottomfraction}{0.85}
\renewcommand{\textfraction}{0.08}
\renewcommand{\floatpagefraction}{0.80}
\setcounter{topnumber}{3}
\setcounter{bottomnumber}{2}
\setcounter{totalnumber}{5}

\shorttitle{ChannelFlow-Tools}

\shortauthors{S.~Kavane et~al.}

\title[mode=title]{ChannelFlow-Tools: A Configuration-Driven Pipeline for Generating Machine-Learning-Ready Datasets of 3D Obstructed Channel Flows}

\author[a]{Shubham Kavane}
\cormark[1]
\ead{shubham.kavane@fau.de}
\credit{Conceptualization, Methodology, Software, Validation, Writing - Original draft}

\author[b]{Lukas Schr\"oder}
\credit{Software, Validation, Writing - Review \& Editing}

\author[a]{Kajol Kulkarni}
\credit{Writing - Review \& Editing}

\author[c]{Fernando Gonzalez}
\credit{Supervision, Writing - Review \& Editing}

\author[a]{Harald K\"ostler}
\credit{Supervision, Funding acquisition, Writing - Review \& Editing}

\affiliation[a]{organization={Chair for Computer Science 10 (System Simulation),
    Friedrich-Alexander-Universit\"at Erlangen-N\"urnberg},
    addressline={Cauerstra\ss{}e 11},
    city={Erlangen},
    postcode={91058},
    country={Germany}}

\affiliation[b]{organization={Erlangen National High Performance Computing Center (NHR@FAU),
    Friedrich-Alexander-Universit\"at Erlangen-N\"urnberg},
    addressline={Martensstra\ss{}e 1},
    city={Erlangen},
    postcode={91058},
    country={Germany}}

\affiliation[c]{organization={CERFACS},
    addressline={42 avenue Gaspard Coriolis},
    postcode={31057},
    city={Toulouse Cedex},
    country={France}}

\cortext[cor1]{Corresponding author}

\begin{abstract}
Data-driven surrogate models are increasingly used in computational fluid dynamics, and their reliability depends on the quality of the training data. These models are typically trained on fixed, pre-generated datasets. Systematic surrogate studies require controlled data generation, in which datasets can be regenerated, adapted, or extended to match specific research requirements. We introduce \textsc{ChannelFlow-Tools}, an open-source, configuration-driven pipeline for generating ML-ready datasets of three-dimensional obstructed channel flows. The pipeline integrates procedural obstacle geometry generation across six shape families, signed-distance-field (SDF) voxelisation, lattice-Boltzmann simulation, and packaging into ML-ready tensors. The workflow is driven by configuration files, with byte-identical reproducibility verified for the geometry-generation stage. The pipeline is evaluated through a full-corpus mesh-integrity audit, analytical and corpus-level validation of the SDF representation, canonical sphere-flow benchmarks for the solver, and a per-scene data-integrity audit. To demonstrate that the pipeline produces physically consistent and directly usable training data, three surrogate models (3D U-Net, FNO, and U-FNO) are trained on a sample dataset of 450 simulations spanning $Re_c \approx 1000$--$10{,}000$, generated entirely through the pipeline. The models learn the geometry-to-flow mapping and show physically interpretable behaviour on shape-family and Reynolds-number out-of-distribution splits, confirming direct downstream usability. \textsc{ChannelFlow-Tools} thus provides shared, auditable infrastructure for controlled benchmarking of geometry-aware CFD surrogates.
\end{abstract}


\maketitle

\section{Introduction}
\label{sec:introduction}

Data-driven surrogate models are increasingly used in computational fluid dynamics to accelerate flow prediction, from academic benchmarks to industrial design studies~\cite{Li2021FNO, Lu2021DeepONet, Kochkov2021PNAS, wen2022_ufno}. Once trained, these models can provide rapid approximations of flow solutions at a fraction of the computational cost of conventional solvers, enabling applications that require repeated or near-real-time evaluations. Their accuracy and reliability, however, depend strongly on the training data, not only on its quantity but also on its coverage of the relevant geometries and regimes, its physical consistency, and the way it was generated. A central unresolved issue is therefore how reliably these models perform when deployment conditions differ from those represented in their training data.

Distribution shift can arise in several forms. Models trained on one class of geometries often perform less accurately on geometries outside that class. Models trained within a limited parameter regime, such as a Reynolds number range, often extrapolate poorly beyond it. Differences in grid resolution between training and deployment can further reduce predictive accuracy~\cite{luo2024cfdbench, xu2023megaflow2d}. Studying these behaviours rigorously requires controlled single-factor experiments, in which one aspect of the data, such as the geometry class, grid resolution or parameter range, is varied while all other elements of the generation process are held fixed. This places two demands on the underlying data generation process. It must be configurable so that a single factor can be varied in isolation, and independently verified so that any observed effect can be attributed to the intended change rather than to an uncontrolled error in the pipeline.

Existing resources meet these two requirements only in part. Open benchmark collections such as PDEBench~\cite{Takamoto2022PDEBench}, CFDBench~\cite{luo2024cfdbench} and The Well~\cite{ohana2024well}, together with large aerodynamic corpora such as DrivAerNet++~\cite{elrefaie_drivaernet++_2024} and AhmedML~\cite{ashton_ahmedml_2024}, have substantially lowered the barrier to surrogate research. Some also release parts of the code used to generate their data. However, many are distributed as fixed corpora in which the flow cases, geometries, resolutions and parameter ranges are predefined by their creators. Changing one factor while holding all others fixed therefore requires access to the original generation workflow. In many cases, this workflow is unavailable or does not include sufficient verification at each stage to independently assess the resulting data. Existing resources address several of these requirements, but the complete
combination remains uncommon, as summarised in
Table~\ref{tab:comparison}.

Computational fluid dynamics is too broad for a single pipeline to cover. We therefore scope this work to three-dimensional obstructed channel flow. This setting is well suited to controlled dataset generation for four reasons. First, its reference physics is well established. Wall-bounded channel flow and canonical bluff-body wakes provide quantitative benchmarks against which the solver protocol can be validated. Second, depending on obstacle geometry and Reynolds number, the flow can exhibit separation, recirculation, vortex shedding and transitions between laminar and turbulent regimes. Third, shape family, geometric parameters, orientation, placement and Reynolds number can be varied explicitly, which supports controlled single-factor studies. Finally, the problem remains computationally tractable at corpus scale while being representative of practical internal flow applications.

We address this need with ChannelFlow-Tools, an open, configuration-driven pipeline for generating ML-ready datasets of three-dimensional obstructed channel flows. The pipeline spans procedural obstacle geometry generation, signed-distance-field (SDF) voxelisation, lattice-Boltzmann simulation, and packaging into co-registered tensors. It is controlled through explicit configuration files and a random seed, while the released inputs and metadata support controlled regeneration and extension. Our contributions are as follows.

\begin{enumerate}
    \item \textbf{Pipeline.} A configuration-driven workflow that produces co-registered, ML-ready tensors from procedural obstacle geometry through SDF construction and lattice-Boltzmann simulation.
    \item \textbf{Stage-wise verification and auditing.} Independent verification of the geometry, SDF, and solver stages, together with a per-scene data-integrity audit of the final packaged tensors.
    \item \textbf{Reproducible release.} An audited obstructed-channel-flow corpus released with its configurations, seeds, and metadata~\cite{channelflowtools_dataset, channelflowtools_software}, with byte-identical reproducibility of the geometry-generation stage verified by SHA-256 hashing of the generated meshes and metadata.
    \item \textbf{Usability demonstration.} Three surrogate baselines (3D U-Net, FNO, and U-FNO) evaluated through a learning-curve study, a shape-family out-of-distribution study, and a Reynolds-number extrapolation study.
\end{enumerate}

We position this work as infrastructure for generating controlled datasets needed to study geometry-aware surrogate modelling. The downstream experiments demonstrate the usability of the generated data and characterise generalisation challenges, while the released pipeline provides a resource for investigating them rather than claiming state-of-the-art surrogate performance.

The remainder of the paper is organised as follows. Section~\ref{sec:related_work} reviews related work on CFD datasets, surrogate modelling, and data-generation workflows. Section~\ref{sec:Pipeline} describes the ChannelFlow-Tools pipeline. Section~\ref{sec:Pipeline_validation} presents per-stage validation, dataset characterisation, and reproducibility analyses. Section~\ref{sec:surrogate_demonstration} reports the downstream surrogate studies, and Section~\ref{sec:conclusion_future_work} concludes with limitations and future work.

\section{Related Work}
\label{sec:related_work}

\paragraph{\textbf{CFD Datasets and Benchmarks.}}
Open dataset releases have lowered the barrier to entry for surrogate research in computational fluid dynamics, and an important difference among them is the level of control they provide over the generated data. Broad PDE benchmarks such as PDEBench~\cite{Takamoto2022PDEBench} and The Well~\cite{ohana2024well} provide large ML-ready datasets for canonical problems, and the Johns Hopkins Turbulence Databases~\cite{Li2008JHTDB, Graham2016ChannelDB} offer high-fidelity DNS reference data, including a turbulent channel-flow database. These resources, however, cover fixed, unobstructed configurations and do not provide parametric obstacle-geometry variation. CFDBench~\cite{luo2024cfdbench} releases mesh-generation and simulation-setup code and varies boundary conditions, physical properties, and geometries, but does not provide a single, independently verified geometry-to-tensor workflow. AirfRANS~\cite{Bonnet2022AirfRANS} demonstrates the value of controlled out-of-distribution evaluation, but it is a fixed corpus limited to a single two-dimensional airfoil family. Automotive datasets such as AhmedML~\cite{ashton_ahmedml_2024}, WindsorML~\cite{ashton_windsorml_2024}, and DrivAerNet++~\cite{elrefaie_drivaernet++_2024} extend this practice to high-fidelity three-dimensional data, but regenerating or extending them requires substantial solver expertise. FlowBench~\cite{tali2024flowbench} provides a large collection of two- and three-dimensional cases over parametric and non-parametric geometries, together with binary masks, signed-distance fields, and case-generation utilities, but does not connect geometry generation, SDF construction, solver validation, and tensor packaging within a single verified and configurable workflow.

\paragraph{\textbf{Geometry Representations and Tensor Alignment.}}
Among geometric encodings for flow surrogates, the signed-distance field (SDF) has become a common choice for representing continuous shape variation. SDF-based representations were used in the CNN model of Guo et al.~\cite{Guo2016CNNFlow}, trained on lattice-Boltzmann data, and were later adopted in DeepCFD~\cite{Ribeiro2021DeepCFD} and related studies. DeepSDF~\cite{Park2019DeepSDF_arXiv} established SDFs as learned implicit shape representations, while GINO~\cite{Li2023GINO} used SDF-conditioned neural operators for large-scale three-dimensional vehicle aerodynamics. Recent benchmarking by Rabeh et al.~\cite{Rabeh2025FlowBench} shows that SDF inputs can outperform binary masks when sufficient data are available. Among the resources compared in Table~\ref{tab:comparison}, geometry-aware input representations are sometimes provided, but dedicated assessments of fidelity to the source geometry and explicit checks of spatial alignment between the geometry and flow fields are generally not reported. Downstream users must therefore usually assume these properties rather than verify them from the released data.

\paragraph{\textbf{Reproducible and Controlled Dataset-Generation Pipelines.}}
Beyond configurability, controlled studies require the generation process itself to be reproducible, and this is not guaranteed by releasing the data alone. Mesnard and Barba~\cite{Mesnard2017Reproducible} showed that fully replicating bluff-body CFD results can be difficult even when open code and data are available, and McConkey et al.~\cite{mcconkey2021curated} similarly note that CFD studies are difficult to reproduce because they depend on many input conditions. Releasing generation code, as several of the datasets above do, is therefore not the same as providing a single configuration-driven workflow that connects geometry generation, simulation, and tensor creation with verification at each stage. APEBench~\cite{Koehler2024APEBench} demonstrates the value of deterministic procedural generation controlled by random seeds, but it is restricted to spectral solvers on periodic domains without geometry. Broader workflow systems such as AiiDA~\cite{Huber2020AiiDA} and Tesseract~\cite{Haefner2025TesseractJOSS} provide reproducible and composable generation in adjacent fields, but do not address obstructed-flow data for surrogate modelling. Table~\ref{tab:comparison} compares the scope and reported generation
capabilities of the closest dataset and workflow resources. Several of these
resources are primarily dataset contributions and are therefore not expected
to provide complete generation workflows.

\begin{table*}[pos=t]
\centering
\footnotesize
\setlength{\tabcolsep}{6pt}
\renewcommand{\arraystretch}{1.3}
\caption{Comparison of generation capabilities across open CFD and PDE resources.
The table compares generation capabilities rather than overall dataset quality. [\yes~provided or reported \quad \part~partial or related \quad \no~not reported]}
\label{tab:comparison}
\begin{tabularx}{\textwidth}{@{}lXccccc@{}}
\toprule
\makecell[bl]{\textbf{Resource}} &
\makecell[bl]{\textbf{Primary contribution}} &
\makecell[b]{\textbf{3D}\\\textbf{parametric}\\\textbf{geometry}} &
\makecell[b]{\textbf{Assessed}\\\textbf{geometry}\\\textbf{representation}} &
\makecell[b]{\textbf{Solver}\\\textbf{benchmark}\\\textbf{validation}} &
\makecell[b]{\textbf{Per-scene}\\\textbf{output}\\\textbf{audit}} &
\makecell[b]{\textbf{Public}\\\textbf{configurable}\\\textbf{workflow}} \\
\midrule
PDEBench
& PDE benchmark with generation code
& \no & \no & \no & \no & \part \\
CFDBench
& CFD dataset with generation scripts
& \no & \no & \no & \no & \part \\
AirfRANS
& Parametric airfoil flow dataset
& \no & \no & \yes & \no & \part \\
FlowBench
& Complex geometry flow dataset
& \yes & \part & \no & \no & \part \\
APEBench
& Procedural PDE benchmark pipeline
& \no & \no & \no & \no & \yes \\
MPLBM-UT
& Porous media LBM workflow
& \part & \yes & \no & \no & \yes \\
\textbf{ChannelFlow-Tools}
& \textbf{Obstructed channel dataset pipeline}
& \yes & \yes & \yes & \yes & \yes \\
\bottomrule
\end{tabularx}

\vspace{3pt}

\end{table*}

As summarised in Table~\ref{tab:comparison}, existing resources provide several capabilities required for controlled surrogate studies, but the full combination remains uncommon. To our knowledge, no existing resource for obstructed channel flow brings together configurable generation of varied three-dimensional obstacle geometries across multiple parametric families, SDF construction, flow simulation, and packaging into co-registered ML-ready tensors, together with separate evaluation of the geometry, SDF, and solver stages and per-scene auditing of the final data. ChannelFlow-Tools integrates these capabilities within a single workflow and releases an audited corpus with the configurations and metadata required for controlled reuse and extension. Byte-identical reproducibility is additionally verified for the geometry-generation stage.

\section{ChannelFlow-Tools Pipeline}
\label{sec:Pipeline}

\begin{figure}
\centering
\includegraphics[width=\textwidth]{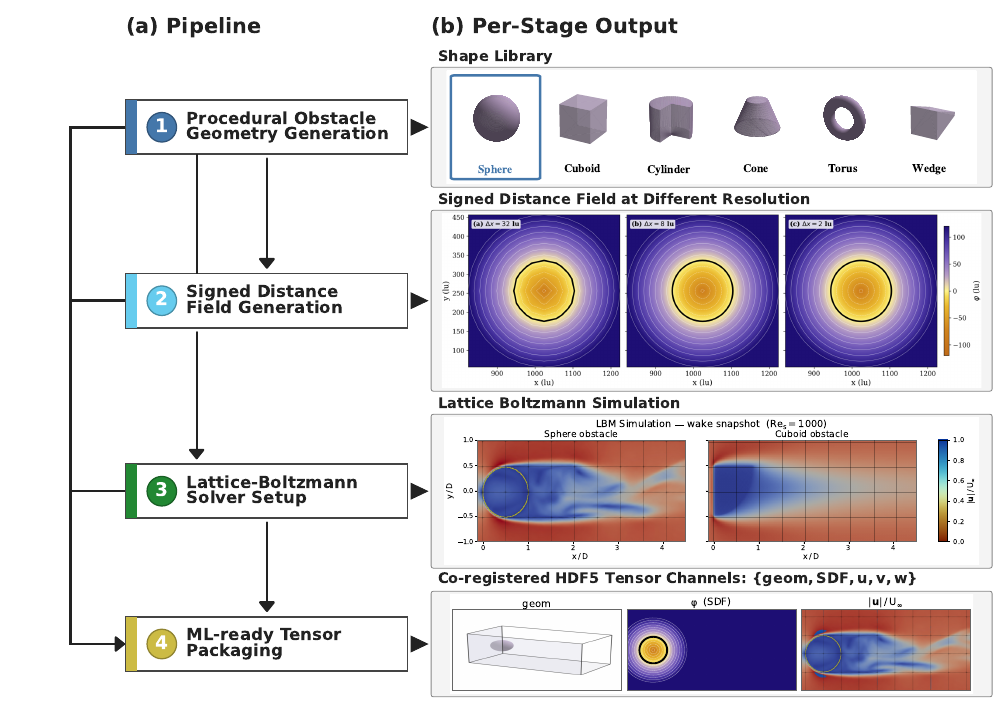}
\caption{Overview of the \textsc{ChannelFlow-Tools} pipeline.
(a)~Four-stage workflow from procedural geometry generation to ML-ready tensor packaging.
(b) Representative per-stage outputs, including obstacle geometries,
signed-distance fields at three voxel spacings
($\Delta x = 32$, $8$, and $2$~lu), a mid-plane visualisation of the
velocity magnitude at $Re_s=1000$, and the final co-registered HDF5 tensor
channels containing the geometry field, SDF $\phi$, and time-averaged
velocity components $(v_x,v_y,v_z)$.}
\label{fig:pipeline_overview}
\end{figure}

This section provides a detailed description of the stages of ChannelFlow-Tools. The first stage is procedural obstacle geometry generation (Section~\ref{sec:geometry}), which produces the three-dimensional obstacle geometry for each scene in the form of an STL (Standard Tessellation Language) file. This file represents the surface as a triangular mesh. Geometry scenes are generated from shape families, geometric parameters, placement constraints, and a random seed specified in the configuration tree. The second stage involves signed-distance-field (SDF) generation (Section~\ref{sec:sdf_generation}), where the generated STL mesh is voxelised onto a Cartesian grid at user-defined resolutions. This produces the geometric representation, or learning grid, that is used as input for the surrogate model. The third and fourth stages are the lattice-Boltzmann solver setup and ML-ready tensor packaging (Section~\ref{sec:solver_lbm}). The lattice-Boltzmann solver stage uses the STL mesh as an obstacle to the flow inside the domain and produces time-averaged and instantaneous velocity fields. The packaging stage, which is integrated into the solver setup, then
resamples the native lattice output onto the same Cartesian grid as the
SDF, producing co-registered HDF5 tensors ready for downstream learning
tasks. The full pipeline is driven by a single Hydra/OmegaConf configuration tree and a single random seed, so that the geometry stage of any released scene can be regenerated deterministically from its configuration alone, and every downstream artefact remains traceable to that configuration.
Figure~\ref{fig:pipeline_overview} summarises the pipeline, and
Table~\ref{tab:pipeline_card} condenses each stage's inputs, outputs, key
configuration parameters, and released artefacts. The key numerical and
configuration parameters required to reproduce the pipeline are summarised
in Appendix Table~\ref{tab:repro-params}, while the complete settings are
provided in the released configuration files. The following subsections
describe each stage in turn.

\begin{table}
\centering
\caption{Pipeline card: the four stages of \textsc{ChannelFlow-Tools}. All stages are
driven by a single Hydra/OmegaConf configuration tree and one random seed; every
artefact carries the same scene identifier, making the path from configuration to
ML-ready tensor traceable.}
\label{tab:pipeline_card}
\small
\begin{tabularx}{\textwidth}{@{}l>{\raggedright\arraybackslash}X>{\raggedright\arraybackslash}X>{\raggedright\arraybackslash}X>{\raggedright\arraybackslash}X@{}}
\toprule
\textbf{Stage} & \textbf{Input} & \textbf{Output} & \textbf{Key configuration} & \textbf{Released artefacts} \\
\midrule
1. Geometry generation &
Configuration tree; random seed; scene index &
STL scene meshes; YAML metadata &
Shape family (6), parameters, object count (1--3), placements, sampling mode &
10{,}060 STL meshes; 15{,}562 objects; YAML records; SHA-256 hash manifests \\
\addlinespace
2. SDF generation &
STL scene mesh; grid specification &
Dense signed-distance field co-registered with the target grid &
Grid resolution and spacing:  $256{\times}128{\times}128$, and   $128{\times}64{\times}64$;  &
350 geometry–SDF verification pairs at $256{\times}128{\times}128$ ($\Delta x = 4$~lu), used for SDF validation \\
\addlinespace
3. LBM simulation &
STL obstacle mesh; flow configuration &
Time-averaged (and instantaneous) velocity fields; force logs &
Domain $1024{\times}512{\times}512$\,lu; $Re_c \approx 100$--$15{,}000$ (supported $Re_c$ range); bounce-back walls &
450 completed flow simulations and corresponding force logs \\
\addlinespace
4. ML tensor packaging &
Native-lattice solver output; SDF grid &
Co-registered HDF5 tensor per scene
(geometry field, SDF $\phi$, and time-averaged velocity components
$v_x$, $v_y$, and $v_z$) &
Target-grid resolution; resampling policy; channel selection &
450 complete simulation scenes, released at both $128{\times}64{\times}64$ ($\Delta x = 8$~lu) and $256{\times}128{\times}128$ ($\Delta x = 4$~lu); \\
\bottomrule
\end{tabularx}
\end{table}

\subsection{Procedural Obstacle Geometry Generation}
\label{sec:geometry}

The geometry generation stage provides the geometric input for the ChannelFlow-Tools pipeline. Its purpose is to create three-dimensional obstacle geometries that define the obstructed channel-flow cases used in the dataset. For each scene,\footnote{A scene refers to one generated obstructed channel-flow configuration, containing one or more obstacle geometries together with their placement and flow parameters.} the obstacle geometry is generated from the configuration tree and exported as an STL mesh. 

The geometry generation stage is fully configuration-driven. All choices required to define a scene, including the shape family, geometric parameters, orientation, placement constraints, number of obstacles, and random seed, are specified in a single Hydra/OmegaConf configuration tree. The random seed controls the sampling process, so that the same configuration can be used to regenerate the same geometry. The geometry tool uses CadQuery ~\cite{cadquery}, a Python CAD scripting library built on OpenCASCADE Technology (OCCT) ~\cite{opencascade}. CadQuery is used to construct the solid obstacle geometries, which are then exported as STL surface meshes for the downstream stages.

The geometry library supports six primitive shape families: sphere, cuboid, cylinder, cone, torus, and wedge. Each family is parameterised by a set of continuous geometric variables, such as radii, lengths, widths, heights, and sweep angles, together with orientation angles and placement coordinates. These families were selected to provide a broad range of geometric features relevant to obstructed channel flows, including smooth curved surfaces, sharp edges, compact bodies, elongated structures, and non-convex shapes. The geometry library can be extended by adding new shape families, each defined through a corresponding CadQuery solid constructor and a set of parameter ranges.

To generate a diverse range of obstacle geometries through variations in geometric parameters, ChannelFlow-Tools supports two parameter-sampling modes. The first is a random mode, in which each parameter is drawn independently from its configured uniform range. The second is a stratified Sobol mode ~\cite{JoeKuoSobol}, in which a low-discrepancy Sobol sequence is mapped across the full input-parameter space, including shape selection, continuous shape parameters, orientation angles, and position coordinates, to obtain quasi-uniform coverage. 

The obstacle geometries are placed carefully inside the channel domain to ensure sufficient upstream, downstream, and wall clearance. Before a candidate scene is accepted, the geometry generation tool applies four feasibility checks:

\begin{enumerate}
\item The axis-aligned bounding box of each obstacle must fit within the channel cross-section while satisfying a minimum wall clearance, $\delta_{\mathrm{wall}}$. 
\item The obstacle centre must lie at least $L_{\mathrm{u}}$ from the inlet and $L_{\mathrm{d}}$ from the outlet. This ensures sufficient flow-development length upstream of the obstacle and helps reduce recirculation artefacts near the outflow boundary.
\item The obstacle volume must exceed a configurable minimum threshold to avoid degenerate cases, such as extremely thin wedges or very small bodies.
\item For multi-object scenes, each pair of obstacles must maintain a minimum inter-object clearance, $\delta_{\mathrm{inter}}$. This prevents overlap, near-contact, or mesh intersection.
\end{enumerate}

\begin{figure}
\centering
\includegraphics[width=0.90\textwidth]{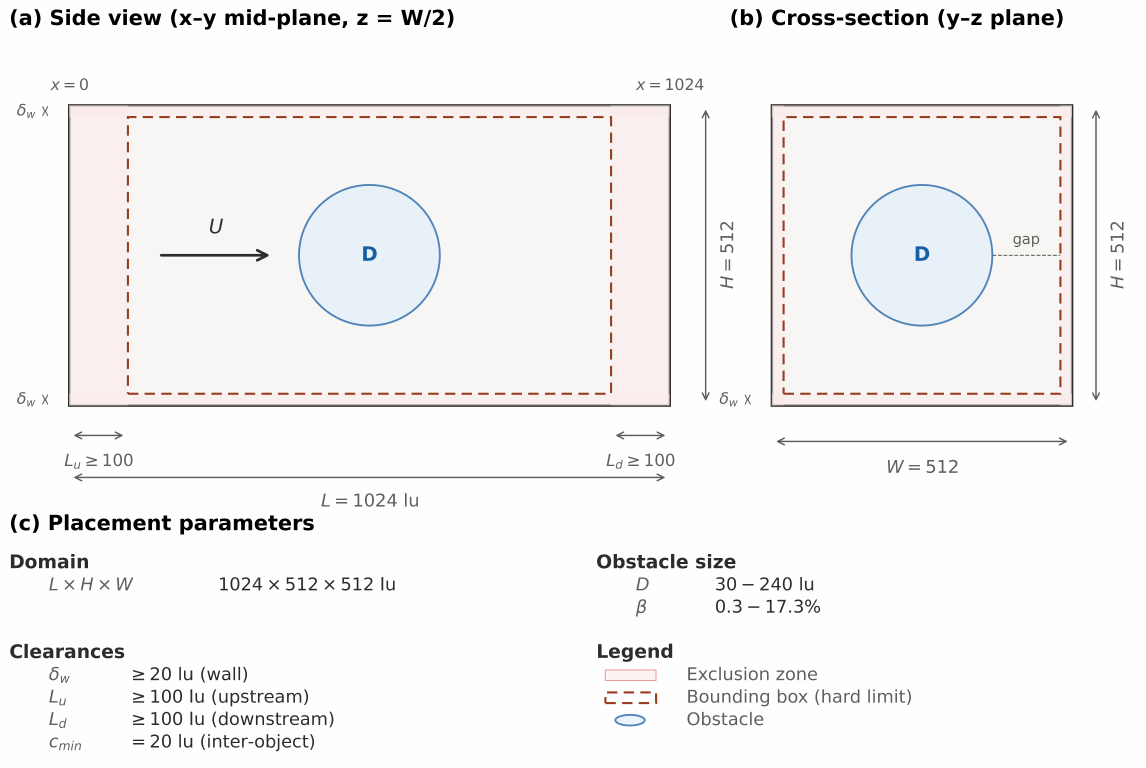}
\caption{Computational domain and obstacle placement constraints.
(a)~Side view ($x$--$y$ mid-plane) showing the channel of length
$L = 1024$~lu and height $H = 512$~lu, with upstream and downstream
exclusion zones $L_{\mathrm{u}}, L_{\mathrm{d}} \geq 100$~lu and
wall clearance $\delta_{\mathrm{wall}}$. The arrow indicates the
bulk flow direction~$U$.
(b)~Cross-section ($y$--$z$ plane) showing the bounding-box hard
limit (dashed red), the centre sampling region (dotted green), and
the minimum gap between the obstacle and channel walls.
(c)~Summary of placement parameters, including obstacle size range
$D = 30$--$240$~lu, blockage ratio $\beta = 0.3$--$17.3\%$, and
clearance thresholds $L_{\mathrm{u}}$, $L_{\mathrm{d}}$,
$\delta_{\mathrm{wall}}$, and $\delta_{\mathrm{inter}}$.}
\label{fig:domain_schematic}
\end{figure}

The domain layout and constraint distances are illustrated in Fig.~\ref{fig:domain_schematic}. Any candidate scene that fails one or more of these checks is discarded and re-sampled until a valid configuration is obtained. Once a candidate scene satisfies the placement and feasibility checks, the final obstacle geometry is exported as a binary STL mesh. Along with the STL file, a YAML metadata record is written for each scene. This metadata stores the shape family, sampled geometric parameters, placement coordinates, orientation, random seed, and relevant mesh-quality statistics. By storing both the mesh and its generation metadata, each scene remains traceable, reproducible, and linked to the configuration from which it was generated.

The exported STL mesh forms the interface between the geometry stage and the rest of the pipeline. In the second stage, the same mesh is converted into a signed-distance field on a Cartesian grid, providing the geometric input representation for the surrogate model. In the third stage, the STL mesh is used again in the lattice-Boltzmann solver setup to define the obstacle geometry inside the flow domain.

\begin{figure}
  \centering
  \includegraphics[width=\textwidth]{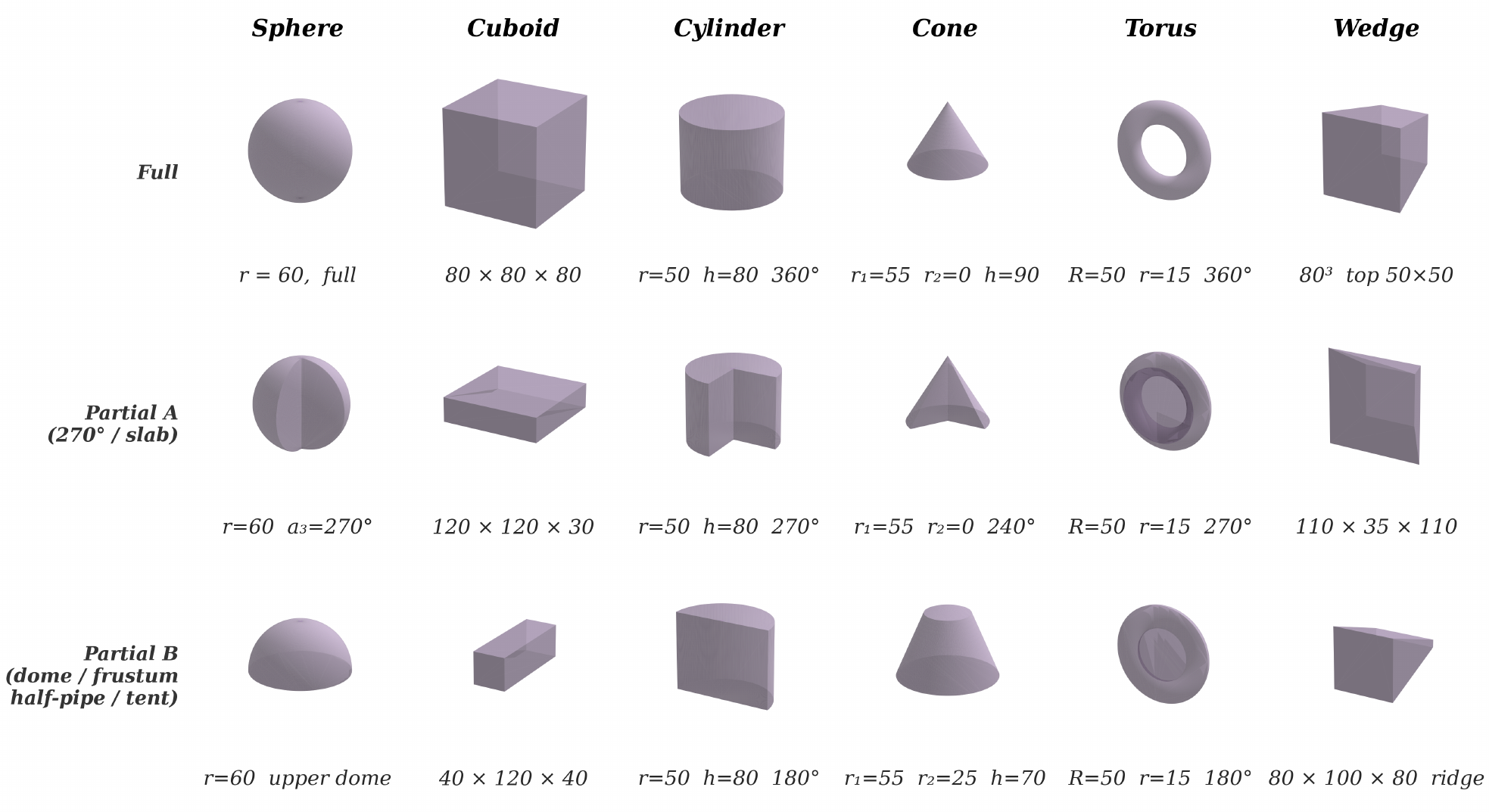}
  \caption{Representative geometries from the six primitive families}
  \label{fig:shape_gallery}
\end{figure}

\subsection{Signed Distance Field Generation}
\label{sec:sdf_generation}

The signed-distance-field (SDF) generation stage converts the STL surface mesh produced by the geometry stage into a structured grid-based representation suitable for convolutional and operator-learning surrogate models. Each exported STL mesh is converted into a voxelised representation on a Cartesian grid to produce a scalar signed-distance field, denoted by $\varphi(\mathbf{x})$. This field provides a compact volumetric representation of the obstacle geometry and serves as the primary geometric input channel for the downstream surrogate-learning task.

The signed-distance field encodes the obstacle geometry as a volumetric scalar field $\varphi(\mathbf{x})$, defined at every point $\mathbf{x}$ in the computational domain. Its magnitude, $|\varphi|$, represents the Euclidean distance to the nearest obstacle surface, and its sign distinguishes the solid interior, $\varphi < 0$, from the fluid exterior, $\varphi > 0$, with $\varphi = 0$ on the obstacle boundary~\citep{OsherFedkiw2003,Sethian1999}. SDF provides a smoother geometric encoding and explicit distance-to-surface information throughout the domain. For an exact SDF, $\lVert \nabla \varphi \rVert = 1$ holds almost everywhere away from non-differentiable medial-axis locations~\citep{Park2019DeepSDF_arXiv}. This makes the SDFs particularly useful for geometry-aware CFD surrogate models, especially near obstacle boundaries where flow features are sensitive to the geometric representation~\citep{Regazzoni2025SDFUSM,Rabeh2025FlowBench}. We therefore adopt the SDF as the primary geometric input for all subsequent surrogate models.

The Cartesian grid used for SDF generation is defined from the domain bounds and voxel spacing specified in the configuration tree. In the present setup, the STL vertex coordinates are used directly in the lattice-unit coordinate system of the simulation domain, $[0,1024] \times [0,512] \times [0,512]$~lu, without any scaling, translation, or mesh repair. The voxel spacing $\Delta x$ is user-configurable, enabling the same STL geometry to be represented at different levels of detail. 

The output of this stage is a dense, three-dimensional SDF array stored alongside the relevant metadata. This metadata ensures that the generated SDF can be traced back to the configuration and geometry used during procedural geometry generation. The same configuration-driven workflow supports multiple SDF resolutions. Based on the convergence and surface reconstruction analysis presented in Section~\ref{sec:sdf_analysis}, the complete simulation corpus is released at two co-registered ML resolutions, $128 \times 64 \times 64$ ($\Delta x = 8$~lu) and $256 \times 128 \times 128$ ($\Delta x = 4$~lu). In addition, a separate set of 350 geometry--SDF pairs is released at $256 \times 128 \times 128$ ($\Delta x = 4$~lu) specifically to reproduce the corpus-scale SDF verification presented in Section~\ref{sec:sdf_corpus}.

The generated SDF field forms the grid-based geometry representation used in downstream surrogate learning. In the solver stage, the same STL mesh is used to define the obstacle geometry within the lattice-Boltzmann simulation. The resulting flow fields are later mapped onto the same Cartesian grid as the SDF. This co-registration ensures that the geometric input and the corresponding flow solution are spatially aligned in the final ML-ready tensors.

\begin{figure}
  \centering
  \includegraphics[width=0.8\textwidth]{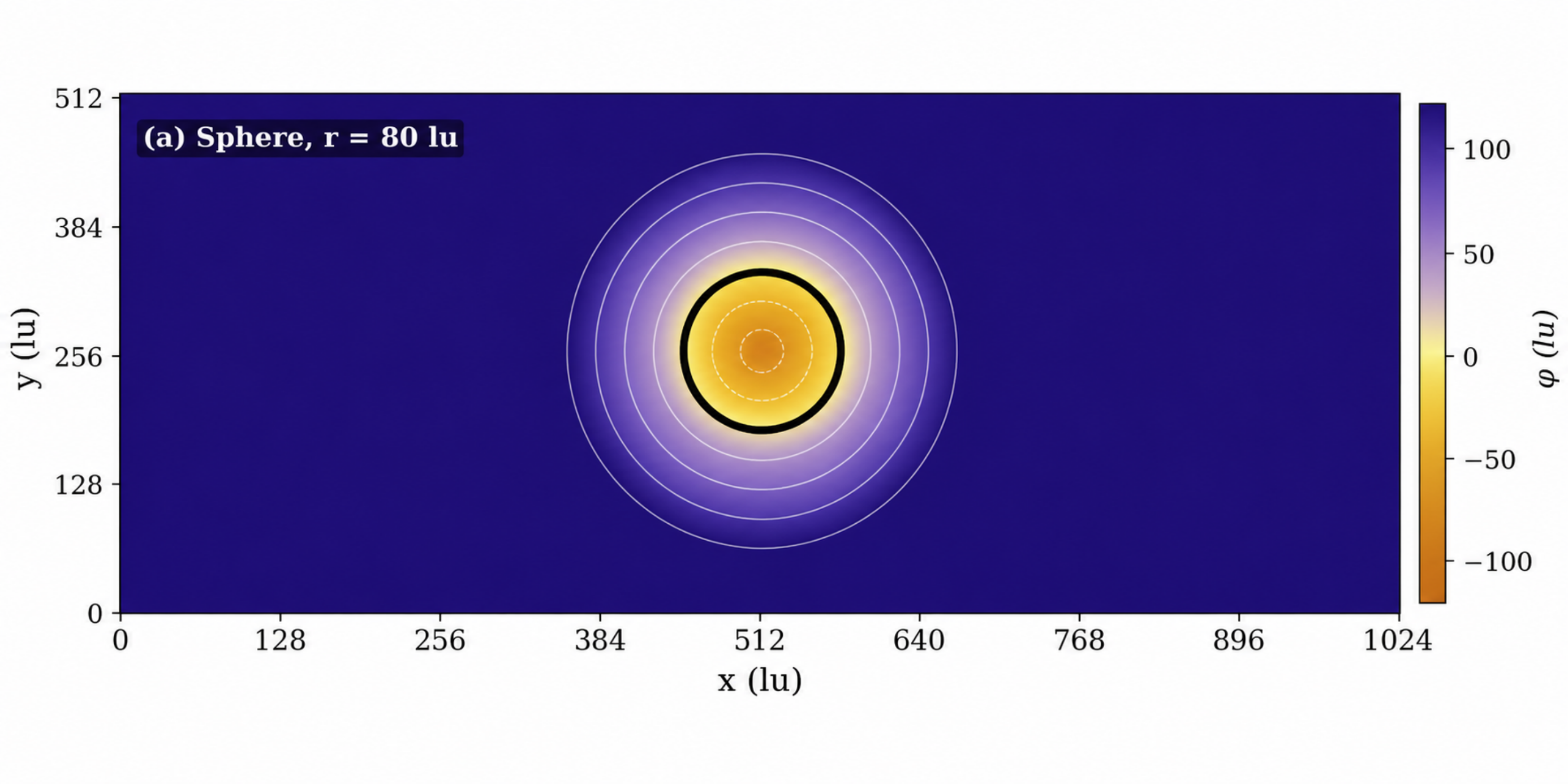}
  \caption{Mid-plane cross-section of the computed SDF for a sphere
    ($r = 80$~lu) over the production channel domain
    $1024 \times 512 \times 512$~lu, voxelised on a $256 \times 128 \times 128$
    grid ($\Delta x = 4$~lu). White lines show isocontours at equal distance
    intervals; the thick black contour marks the zero-level-set
    ($\varphi = 0$).}
  \label{fig:sdf_cross_section}
\end{figure}

\subsection{Lattice-Boltzmann Solver Setup}
\label{sec:solver_lbm}

The lattice-Boltzmann solver setup constitutes the simulation stage of the ChannelFlow-Tools pipeline. Once the obstacle geometry has been generated and converted into an SDF representation, the same STL mesh is used to define the solid obstacle within the flow domain. The purpose of this stage is to compute the reference flow solution corresponding to each generated geometry. These simulated velocity fields then provide the target data for training the surrogate model. ChannelFlow-Tools does not introduce a new CFD solver; instead, it integrates the open-source \textsc{waLBerla} framework into the configuration-driven dataset-generation workflow~\cite{Bauer2020_walberla_camwa}. The solver protocol follows the setup introduced and verified in the companion work by Schr\"oder and Kavane et al.~\cite{Schroeder2025LBMDataset}, while the present pipeline focuses on integrating it across generated scenes and extending the verification setup to per-simulation outputs.

The solver is based on the lattice Boltzmann method using a D3Q27 velocity stencil and a cumulant collision operator~\cite{geier2015cumulant}, combined with a fourth-order advection--diffusion correction~\cite{geier2017parametrization} for cases with $Re_c \geq 5000$. This provides stable simulations of three-dimensional obstructed channel flows over a wide range of Reynolds numbers. All simulations are performed in lattice units~\cite{Kruger2017LBM} on a domain of size $1024 \times 512 \times 512$ in the streamwise, wall-normal, and spanwise directions, respectively. This domain matches the coordinate system used for geometry and SDF generation. The channel height is used as the reference length scale for defining the Reynolds number. The STL mesh generated in the geometry stage is inserted into the channel as the solid obstacle to obstruct the flow. A uniform velocity boundary condition is imposed at the inlet, and an
extrapolation outflow condition (\texttt{ExtrapolationOutflow}) is used at
the outlet. No-slip conditions are
enforced using linear bounce-back at the channel walls and quadratic
bounce-back at the obstacle surface~\cite{geier2015cumulant}.

Each simulation runs for a total of twenty flow-through times. One
flow-through time is defined as $L_x/U_\infty$, corresponding to the number of
lattice time steps required for the inlet flow to traverse the full
streamwise domain length. For $L_x=1024$ lattice cells and
$U_\infty=0.05$, one flow-through time corresponds to approximately
$20{,}480$ lattice time steps. The first fifteen flow-through times serve
as the warm-up phase, during which start-up transients dissipate and the
wake develops around the obstacle. Time-averaged velocity fields are then
accumulated over the final five flow-through times and form the main target
quantities for surrogate model training. The solver additionally supports
the output of instantaneous flow fields for diagnostic purposes and future
time-dependent surrogate models. Forces acting on the obstacle are
integrated during the averaging window and written to a force log, from
which drag and lift coefficients can be derived during post-processing.

The fourth stage of ChannelFlow-Tools produces ML-ready tensors and is integrated into the solver setup. During this process, the simulation output is resampled from the native lattice grid to a user-defined structured target grid and stored as a single HDF5 file for each simulation. This step co-registers the resampled flow fields with the SDF generated in the previous stage on a common Cartesian grid. This creates an ML-ready input--target pair, aligning the geometric representation and the corresponding flow solution spatially. Each packaged sample is linked to the same scene identifier used throughout the geometry, SDF, and solver stages. This makes the complete path from configuration to STL mesh, SDF field, CFD solution, and ML-ready tensor traceable. The complete simulation corpus is packaged and released at both target grids, $128 \times 64 \times 64$ ($\Delta x = 8$~lu) and $256 \times 128 \times 128$ ($\Delta x = 4$~lu).

Appendix Table~\ref{tab:dataset_card} summarises the released corpus and its access details.

\section{Pipeline Validation}
\label{sec:Pipeline_validation}

The value of a data-generation pipeline rests on whether its outputs can be
trusted without re-verification by every downstream user. We therefore validate
\textsc{ChannelFlow-Tools} at three levels: correctness of each individual stage
(geometry, SDF, solver), statistical characterisation of the corpus that the
stages produce at scale, and usability of the final tensors for supervised
learning. Table~\ref{tab:validation_map} maps each claim made about the pipeline
to the corresponding validation evidence; the following subsections present each
experiment in turn.

\begin{table}[pos=t]
\centering
\caption{Validation map: each claim about \textsc{ChannelFlow-Tools} and the evidence supporting it.}
\label{tab:validation_map}
\small
\begin{tabularx}{\columnwidth}{@{}>{\raggedright\arraybackslash}X>{\raggedright\arraybackslash}Xl@{}}
\toprule
\textbf{Claim} & \textbf{Evidence} & \textbf{Section} \\
\midrule
Geometry generation is robust &
Mesh-integrity audit of the full corpus; STL-to-SDF bridge fidelity &
\S\ref{sec:v1_mesh_integrity}--\ref{sec:v2_sdf_bridge} \\
\addlinespace
Sampling covers the configured design space &
Parameter-space and geometric-feature-space coverage characterisation &
\S\ref{sec:v3-design-space-coverage} \\
\addlinespace
Mesh quality does not degrade with corpus size &
Descriptive Kolmogorov--Smirnov distance comparison &
\S\ref{sec:v5_scale_invariance} \\
\addlinespace
Upstream generation is reproducible &
SHA-256 byte-identical regeneration from configuration, seed, and scene index &
\S\ref{sec:v6_reproducibility} \\
\addlinespace
SDFs are numerically and geometrically accurate &
Analytical-primitive convergence, zero-isosurface reconstruction, corpus-scale audit against source STL &
\S\ref{sec:sdf_analysis} \\
\addlinespace
The LBM solver protocol is valid &
Canonical sphere-flow benchmark: wake statistics, vortex shedding, drag-coefficient comparison &
\S\ref{sec:solver_validation_sphere} \\
\addlinespace
Released flow tensors are physically consistent &
Per-scene data-integrity audit (completeness and stability, mass conservation, averaging convergence, geometry–flow registration) &
\S\ref{sec:solver_data_integrity} \\
\addlinespace
The dataset is usable for ML &
3D U-Net, FNO, and U-FNO baselines trained on the released tensors, including OOD generalisation splits &
\S\ref{sec:surrogate_demonstration} \\
\bottomrule
\end{tabularx}
\end{table}

\subsection{Obstacle Geometry Tool Validation}
\label{sec:geom_validation}
 
The geometry creation tool is validated through five experiments that
progressively establish mesh integrity at the individual-mesh level,
fidelity of the interface to the downstream SDF stage, stability of mesh
quality at production scale, determinism of the generation process, and
completeness of design-space coverage.

\subsubsection{Mesh Integrity}
\label{sec:v1_mesh_integrity}

The geometry generation stage must produce STL meshes that can be passed to the downstream SDF and simulation workflow without the need for manual repair. We therefore audited the generated obstacle geometries, consisting of 6,036 single-object scenes, 2,546 two-object scenes, and 1,478 three-object scenes. As each scene is exported as a single scene-level STL file, a total of 10,060 STL meshes are produced.

Table~\ref{tab:mesh_integrity_audit} summarises the seven mesh-quality checks used in the audit, what each check measures, and the corresponding result. The main geometric validity checks were passed across the obstacle geometries: all scenes have positive volume, the expected number of connected components, and remain within the computational domain. The Euler characteristic check also passed for all single-object scenes, where a unique reference topology can be defined.

The only raw failures occur in watertightness, degenerate triangle, and minimum edge checks. These failures occur in the same 1,178 scenes and can be traced back to a local OpenCASCADE tessellation artefact at parametric singularities, mainly at sphere poles and cone apices. These artefacts do not correspond to invalid CAD solids, misplaced objects, disconnected components, or macroscopic holes in the generated geometry. The artefact is most prevalent in the parametric sweep variants of spheres, where the tessellation can generate extremely thin triangles near pole singularities. Although this can be mitigated by restricting the obstacle size range, the present corpus deliberately retains small sizes in order to evaluate the geometry stage thoroughly.

The downstream relevance of these mesh-level artefacts is assessed through
the STL-to-SDF bridge-fidelity experiment. The experiment uses a
preselected stratified subset of 350 scenes covering all six shape families
and multi-object configurations. All 350 selected scenes completed SDF
generation and zero-level-surface reconstruction successfully. Across this
subset, the reconstructed surfaces retain sub-voxel mean accuracy. These
results show that the local tessellation artefacts do not prevent accurate
SDF generation for the evaluated cases or require manual repair before
downstream processing.

\begin{table*}[pos=t]
  \centering
  \caption{Mesh-integrity audit of the production geometry corpus
  ($N = 10{,}060$ scene-level STL meshes).}
  \label{tab:mesh_integrity_audit}

  \begin{tabular}{lll}
    \toprule
    Criterion & Purpose & Result \\
    \midrule
    Watertightness & Closed solid--fluid boundary & $8{,}882/10{,}060$ -- pass \\
    Positive volume & Non-zero enclosed volume & $10{,}060/10{,}060$ -- pass \\
    Expected components & Expected object parts & $10{,}060/10{,}060$ -- pass \\
    Euler characteristic & Expected mesh topology & $6{,}036/6{,}036$ -- pass \\
    Degenerate triangles & No near-zero-area facets & $8{,}882/10{,}060$ -- pass \\
    Domain containment & Obstacle inside channel & $10{,}060/10{,}060$ -- pass \\
    Minimum edge length & No extremely small edges & $8{,}882/10{,}060$ -- pass \\
    \bottomrule
  \end{tabular}
\end{table*}

\subsubsection{STL-to-SDF Bridge Fidelity}
\label{sec:v2_sdf_bridge}

The STL-to-SDF bridge-fidelity experiment evaluates whether the surface
geometry produced by the geometry stage is preserved in the downstream SDF representation. We evaluate this STL-to-SDF interface on a stratified verification subset of $350$ scenes rather than on the full $10{,}060$-scene geometry corpus, since this test requires SDF generation, zero-level-surface extraction, and surface-distance comparison for each case. The subset comprises $50$ scenes from each of the six single-object shape families, as well as $50$ multi-object scenes, consisting of $36$ two-object and $14$ three-object configurations. This design provides balanced coverage of the supported shape families while also including the more geometrically complex multi-object cases.

Three surface-distance metrics are used to quantify the reconstruction error. The Hausdorff distance ($H$) measures the largest local deviation between the reconstructed SDF surface and the original STL surface. It therefore captures the worst-case mismatch. The mean surface distance (MSD) measures the average deviation between the two surfaces and is less sensitive to isolated outlier points. The Chamfer distance provides a symmetric, nearest-neighbour measure of surface mismatch by comparing the two surfaces in both directions. Together, these metrics measure both the average reconstruction quality and the largest local deviations.

Reconstruction is considered successful when the resolution-normalised mean surface distance satisfies $\mathrm{MSD}/\Delta x \leq 1.0$. This requires the reconstructed surface to lie within one voxel of the original STL surface on average and serves as the acceptance criterion used to calculate the reported pass rates. This is a natural tolerance for a grid-based SDF representation because geometric details below the grid spacing cannot be consistently resolved in the sampled field. The normalised Hausdorff distance, $H/\Delta x$, is reported separately as a secondary worst-case diagnostic. A reference value of $H/\Delta x \leq 5.0$ is used to identify large local deviations, but it does not determine the reconstruction pass rate because the Hausdorff distance is controlled by a single worst point and can be affected by isolated sharp features, close object gaps, or parametric singularities even when the reconstructed surface remains accurate on average.

As summarised in Table~\ref{tab:bridge_fidelity}, all six single-object families achieve sub-voxel mean surface accuracy and pass at $100\%$. Multi-object scenes exhibit larger worst-case deviations due to closer object spacing and increased geometric complexity; however, their mean surface distance remains sub-voxel. These results demonstrate that the SDF zero-level surface remains well aligned with the original STL geometry, including the sphere cases affected by local tessellation artefacts identified in the mesh-integrity audit.

\begin{table*}[pos=t]
  \centering
    \caption{STL-to-SDF round-trip fidelity per stratum
    (grid spacing $\Delta x=4$~lu, $256\times128\times128$ grid).
    The pass rate is calculated using the acceptance criterion
    $\mathrm{MSD}/\Delta x\leq1.0$; the Hausdorff distance is reported as a
    secondary worst-case diagnostic.}
  \label{tab:bridge_fidelity}
  \begin{tabular}{lrrrrr}
    \toprule
    Stratum & $N$ & Mean $H/\Delta x$ & Max $H/\Delta x$
            & $\mathrm{MSD}/\Delta x$ & Pass rate \\
    \midrule
    Cuboid               & 50 & 0.98 &  1.47 & 0.27 & 100.0\% \\
    Sphere               & 50 & 1.04 &  4.44 & 0.23 & 100.0\% \\
    Torus                & 50 & 1.10 &  3.29 & 0.29 & 100.0\% \\
    Cylinder             & 50 & 1.15 &  5.78 & 0.25 & 100.0\% \\
    Wedge                & 50 & 1.16 &  2.96 & 0.23 & 100.0\% \\
    Cone                 & 50 & 1.35 &  7.04 & 0.25 & 100.0\% \\
    Multi-object (2-obj) & 36 & 1.82 &  8.51 & 0.35 &  94.4\% \\
    Multi-object (3-obj) & 14 & 3.99 & 18.65 & 0.82 &  92.9\% \\
    \bottomrule
  \end{tabular}
\end{table*}

\subsubsection{Scale Stability of Mesh Quality}
\label{sec:v5_scale_invariance}

The geometry generation pipeline should maintain consistent mesh quality as
the size of the corpus increases. To assess this, we compared four nested
subsets of the production corpus containing $50$, $500$, $5{,}000$, and
$10{,}000$ scenes. Each subset was taken as a prefix of the same Sobol
sequence, representing successive stages in the growth of the corpus.

Three mesh quality metrics were examined. Triangle count measures the
complexity of the exported surface mesh and is related to storage and
processing costs. Surface area describes the geometric scale and exposed
obstacle surface. Minimum edge length measures local tessellation quality
and indicates whether extremely small mesh elements become more common as
the corpus grows.

For each pair of corpus sizes, we calculated the Kolmogorov--Smirnov
distance $D$ between the corresponding empirical distributions. The value
of $D$ is the largest absolute difference between two empirical cumulative
distribution functions. Smaller values therefore indicate more similar
distributions. Because the subsets are nested and share scenes, the
comparisons are not independent. We consequently use $D$ only as a
descriptive measure of distributional stability and do not report
inferential $p$ values.

Table~\ref{tab:ks_tests} reports the resulting distances. For triangle
count, $D$ ranges from $0.011$ to $0.103$. For surface area, the largest
difference is $D=0.152$ for the comparison between the $50$ and $500$ scene
subsets, while the remaining values are at or below $0.116$. For minimum
edge length, all values are at or below $0.086$.

The distances generally decrease as the compared subsets become larger.
The smallest values occur between the $5{,}000$ and $10{,}000$ scene
subsets, where $D$ is approximately $0.01$ for all three metrics. Overall,
the empirical distributions remain similar as the Sobol sequence is
extended, indicating that corpus growth does not produce a substantial
descriptive change in the examined mesh quality characteristics.

\begin{table*}[pos=t]
  \centering
  \caption{Kolmogorov--Smirnov distances between the empirical mesh quality
  distributions at different corpus sizes. The distances are reported as
  descriptive measures because the compared subsets are nested.}
  \label{tab:ks_tests}
  \footnotesize
  \setlength{\tabcolsep}{8pt}
  \begin{tabular}{lccc}
    \toprule
    Scale pair
    & Triangle count
    & Surface area
    & Min.\ edge length \\
    \midrule
    $50$ vs.\ $500$
      & $0.082$
      & $0.152$
      & $0.082$ \\
    $50$ vs.\ $5{,}000$
      & $0.103$
      & $0.115$
      & $0.086$ \\
    $50$ vs.\ $10{,}000$
      & $0.097$
      & $0.116$
      & $0.086$ \\
    $500$ vs.\ $5{,}000$
      & $0.047$
      & $0.042$
      & $0.047$ \\
    $500$ vs.\ $10{,}000$
      & $0.042$
      & $0.041$
      & $0.043$ \\
    $5{,}000$ vs.\ $10{,}000$
      & $0.011$
      & $0.012$
      & $0.012$ \\
    \bottomrule
  \end{tabular}
\end{table*}

\subsubsection{Seeded Reproducibility}
\label{sec:v6_reproducibility}

A central design requirement of the geometry pipeline is that any scene can be regenerated from its configuration file, seed, input files, and scene index. We verify this requirement by testing bitwise reproducibility of the generated STL meshes and YAML metadata files under fixed configurations. The test covers both sampling modes supported by the pipeline: stratified Sobol sampling, used for the production corpus, and pseudo-random sampling.

For both Sobol and pseudo-random sampling modes, the corpus-generation experiment was repeated using the same configuration, seed, input files, and sampling settings. In the Sobol case, the archived Sobol sampling schedule was also reused. For each mode, the STL mesh and YAML metadata file produced for every scene in the first run were compared against the corresponding files from the second run using SHA--256 hashes. All corresponding files matched bit-for-bit, showing that the geometry corpus is exactly reproducible under the pinned generation environment for both supported sampling paradigms.

To support independent verification, the public release will include the configuration files, seed values, input-file references, scene indices, Sobol sampling schedule, SHA--256 hash manifests, and verification scripts. A lightweight $1{,}000$-scene reproducibility example is also provided so that users can regenerate a representative subset and compare the resulting STL and YAML hashes against the released reference hashes without rebuilding the full corpus.

These results show that reproducibility is a property of the full geometry-generation pipeline, not of a single sampling method. Under the pinned release environment, a scene can be regenerated and verified from its configuration file, seed, input files, and scene index. 
\subsubsection{Design-Space Coverage}
\label{sec:v3-design-space-coverage}

Design-space coverage is assessed at two levels. The first is the configured
input-parameter space from which the geometries are sampled. The second is
the realised geometric-feature space of the generated meshes. Broad
input-parameter coverage does not necessarily produce broad geometric
diversity because different parameter combinations can result in similar
shapes. We therefore evaluate both levels.

For the input-parameter space, the production corpus contains $10{,}060$
scenes and $15{,}562$ individual objects generated using a stratified Sobol
sampler over $206$ joint input dimensions. These dimensions include shape
selection, continuous geometric parameters, orientation, and position. All
$83$ possible unordered one-, two-, and three-object shape compositions are
represented, showing that the feasibility filter does not remove any
composition class. Most continuous shape parameters cover at least
$99.5\%$ of their configured ranges. The few cases below $95\%$ are mainly
associated with sphere orientation parameters, which are inactive because
spheres are rotationally invariant, and the second torus sweep angle.

\begin{figure}[pos=htbp]
\centering
\includegraphics[width=0.75\textwidth]{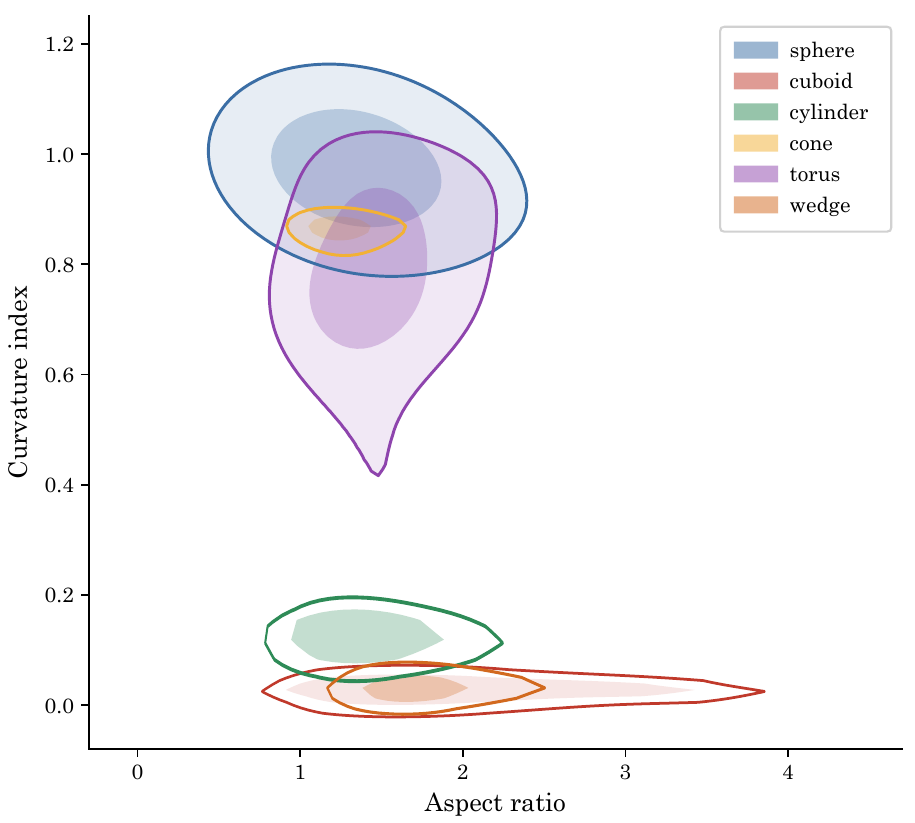}
\caption{Geometric feature-space coverage of the generated single-object corpus in the aspect-ratio--curvature-index plane. Shaded regions are per-family KDE density contours at 25\% (light) and 75\% (dark) of each family's peak density.}
\label{fig:feature-space-coverage}
\end{figure}

For the realised geometric-feature space, we characterise the generated
meshes using aspect ratio, curvature index, sphericity, and convexity to
determine whether the sampled parameters produce geometrically diverse
obstacles. Figure~\ref{fig:feature-space-coverage} shows the six
single-object families in the aspect-ratio--curvature-index plane. The
families occupy distinct but partially overlapping regions. Spheres,
cones, and tori are concentrated in the higher-curvature region, while
cuboids, cylinders, and wedges remain closer to the lower-curvature
region. Aspect ratio further separates compact geometries from elongated
ones. Some overlap is expected because different families can share
similar geometric features, such as the curved side walls and planar caps
of cylinders.

Across the corpus, the realised geometries span $98.4\%$ of the bounded
sphericity range, $99.6\%$ of the bounded convexity range, and the complete
bounded curvature-index range. Pairwise Kolmogorov--Smirnov tests further
show that every shape family differs from every other family in at least
one geometric descriptor at $p<10^{-15}$.

Together, these results show that the Sobol sampler and feasibility filter
preserve broad coverage of the configured input parameters and produce a
diverse realised geometry corpus with distinguishable regions for the
different shape families.

\subsection{Signed-distance-field validation}
\label{sec:sdf_analysis}

The signed distance field (SDF) is the geometric input used by the downstream surrogate models. We therefore validate whether it correctly represents both the sampled distance-field values and the reconstructed obstacle boundary. The validation is organised into three parts.

First, Section~\ref{sec:sdf_convergence} evaluates field-value convergence for analytical geometries whose exact signed-distance functions are known: sphere, cylinder, and axis-aligned cuboid. We compare the OpenVDB-generated SDF against closed-form signed-distance functions over five grid spacings.

Second, Section~\ref{sec:sdf_surface_recon} evaluates the accuracy of the reconstructed zero-isosurface for the same analytical primitives. This complements the field-value test by checking whether the extracted obstacle boundary remains geometrically aligned with the reference surface. 

Third, Section~\ref{sec:sdf_corpus} extends the validation to production geometries from the released corpus, including all six shape families and two- and three-object scenes. Since these geometries do not all admit simple closed-form SDFs, we compare the reconstructed zero-level surface directly against the source STL mesh. 
The main surface-fidelity criterion is that the mean surface distance remains below one voxel, i.e. $\mathrm{MSD}/\Delta x \leq 1.0$. The Hausdorff distance is reported as a secondary worst-case diagnostic, against a reference value of $H/\Delta x \leq 5$, because it is more sensitive to isolated sharp-feature points and local reconstruction artefacts than the mean surface distance.

\subsubsection{Field-value convergence against analytical SDFs}
\label{sec:sdf_convergence}

The field-value validation follows the Method of Manufactured Solutions (MMS)~\citep{Roache1997,Oberkampf2010}: for geometries with known analytical signed-distance functions, the generated SDF should approach the exact reference under grid refinement. This provides a direct verification of the OpenVDB voxelisation stage, independent of any downstream solver quantity.

We use three primitive families for this test: sphere, cylinder, and axis-aligned cuboid. The sphere tests a smooth curved boundary, the cylinder adds sharp circular cap edges, and the cuboid provides a planar control case whose STL surface is represented exactly up to floating-point precision. All primitives are generated with the same geometry and tessellation settings used in the production corpus and are voxelised over five grid spacings, $\Delta x \in \{16,8,4,2,1\}$~lu. The cell-wise $L_2$ error is then computed against the corresponding analytical SDF.

\begin{figure}[pos=htbp]
\centering
    \includegraphics[width=\linewidth]{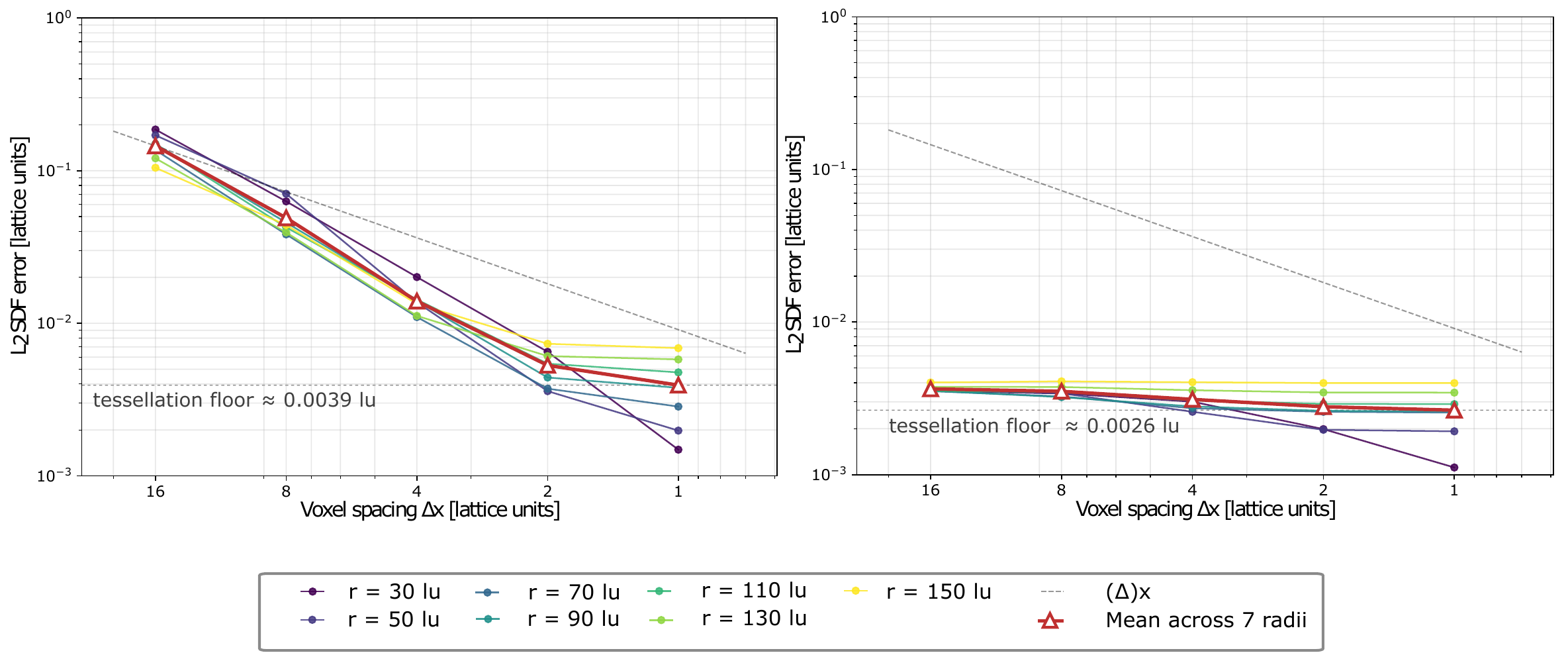}
  \caption{$L_2$ convergence of the OpenVDB-generated SDF against analytical references for sphere and cylinder primitives over $\Delta x \in \{16,8,4,2,1\}$~lu. The sphere shows resolved convergence with fitted rate $\alpha = 1.69 \pm 0.12$ before reaching the tessellation floor, while the cylinder is already close to the tessellation-controlled asymptote across the tested resolutions.}

  \label{fig:sdf_convergence}
\end{figure}

Figure~\ref{fig:sdf_convergence} shows the convergence behaviour for the sphere and cylinder families. For spheres, the $L_2$ error decreases consistently over the resolved regime, with fitted convergence rate $\alpha = 1.69 \pm 0.12$. At finer resolutions, the curves approach a stable error floor, indicating that the remaining discrepancy is no longer controlled by voxel spacing but by the finite STL tessellation of the curved surface.

The cylinder results show the same error mechanism from a different direction. The $L_2$ error is already close to the tessellation-controlled floor across the tested resolutions, so additional voxel refinement changes the error only weakly. Thus, for the configured geometry tolerance, the cylinder SDF is already grid-resolved at the tested spacings.

The cuboid control confirms this interpretation. Since its planar faces are exactly represented by the STL mesh, the SDF error remains close to numerical round-off and does not vary meaningfully with grid spacing. This demonstrates that OpenVDB's distance evaluation is accurate when the input geometry is exact, and that the residual error for curved primitives originates mainly from the geometry tessellation.

At both simulation release tiers, the field-value errors remain small compared with the voxel size. The SDF stage is therefore sufficiently resolved for the configured production geometry, and further accuracy gains would primarily require tighter STL tessellation rather than only a finer SDF grid.

\subsubsection{Zero-isosurface reconstruction accuracy}
\label{sec:sdf_surface_recon}

The convergence study in Section~\ref{sec:sdf_convergence} verifies the sampled SDF values over the full domain. For downstream surrogate models, however, the most relevant geometric information is concentrated near the obstacle boundary. The zero-isosurface of the SDF defines the obstacle location and determines the wall-distance information seen by the model. We therefore also evaluate how accurately the extracted zero-level surface reconstructs the reference obstacle boundary.

For the same sphere, cylinder, and cuboid cases, the zero-isosurface is extracted using marching cubes~\citep{Lorensen1987} and compared with the corresponding reference surface. The main quantity reported here is the mean surface distance (MSD), because it measures the typical boundary displacement and is less sensitive to isolated outlier vertices than worst-case measures. We also report the voxel-normalised value $\mathrm{MSD}/\Delta x$, which expresses the same error relative to the grid spacing.

Table~\ref{tab:sdf_surface_recon_release} summarises the reconstruction accuracy at the simulation resolution $256 \times 128 \times 128$, $\Delta x = 4$~lu. The reconstructed surfaces remain well below one voxel from the reference boundary on average: the MSD is $1.60$~lu for the sphere, $1.35$~lu for the cuboid, and $2.00$~lu for the cylinder, corresponding to $\mathrm{MSD}/\Delta x = 0.400$, $0.338$, and $0.500$, respectively. Thus, the extracted zero-isosurfaces lie, on average, within approximately half a voxel of the analytical surface. This satisfies the sub-voxel criterion used later for the production-geometry audit in Section~\ref{sec:sdf_corpus}.

The relative ordering of the three primitives is consistent with their geometry. The cuboid gives the smallest reconstruction error because its planar faces are represented exactly by the STL mesh. The sphere introduces curved-surface tessellation effects, while the cylinder contains both curvature and sharp cap-edge features, producing the largest, but still sub-voxel, mean deviation at this tier. The full resolution ladder is reported in Table~\ref{tab:sdf_surface_recon_appendix} in \ref{app:sdf}, and the corresponding SDF cross-sections are shown in Fig.~\ref{fig:sphere_multiresolution} in \ref{app:sdf}. These appendix results show that the absolute MSD rapidly reaches a stable plateau, while the zero-level contour transitions from a visibly faceted coarse-grid reconstruction to a smooth boundary at finer resolutions. At resolutions finer than this tier, $\mathrm{MSD}/\Delta x$ can increase because the voxel spacing decreases below this absolute reconstruction floor; this reflects the normalisation by $\Delta x$, not a degradation of the physical surface accuracy.

Overall, the surface-reconstruction analysis confirms that the generated SDFs are not only accurate as sampled distance fields, but also preserve the obstacle boundary with sub-voxel average accuracy at the $\Delta x = 4$~lu simulation resolution, $256 \times 128 \times 128$.

\begin{table}[pos=t]
\centering
\caption{Surface-reconstruction accuracy at the simulation resolution $256\times128\times128$, $\Delta x = 4$~lu. MSD is reported in lattice units and normalised by the voxel spacing.}
\label{tab:sdf_surface_recon_release}
\begin{tabular}{l c c c}
\toprule
Primitive & $\Delta x$ (lu) & MSD (lu) & MSD$/\Delta x$ \\
\midrule
Sphere   & $4$ & $1.60$ & $0.400$ \\
Cuboid   & $4$ & $1.35$ & $0.338$ \\
Cylinder & $4$ & $2.00$ & $0.500$ \\
\bottomrule
\end{tabular}
\end{table}

\subsubsection{Corpus-scale SDF audit}
\label{sec:sdf_corpus}

The analytical tests in Sections~\ref{sec:sdf_convergence} and~\ref{sec:sdf_surface_recon} validate the SDF stage on primitive geometries with closed-form reference fields. To extend this validation to production geometries, we use a separate SDF verification subset of $350$ geometry--SDF pairs selected from the 10,060-scene geometry corpus. This subset is distinct from the 450 complete simulation scenes and contains no flow fields; it is released at $256 \times 128 \times 128$ ($\Delta x = 4$~lu) and is provided specifically to reproduce the corpus-scale SDF validation reported here. The subset covers all six single-object families together with two-object and three-object configurations.

The audit is performed on this preselected stratified subset of
$350$ scenes. All selected scenes completed
SDF generation and zero-level-surface reconstruction successfully. For
each scene, the reconstructed zero-level surface is compared with the
corresponding source STL mesh. Surface fidelity is assessed using the
acceptance criterion $\mathrm{MSD}/\Delta x \leq 1$. The normalised
Hausdorff distance, $H/\Delta x$, is reported separately as a secondary
worst-case diagnostic, as described above.

We also evaluate two aggregate field-level properties. Sign consistency
measures the percentage of sampled points whose SDF sign agrees with an
independent inside--outside classification. The eikonal residual measures
the mean deviation of $\|\nabla\varphi\|$ from unity within the evaluated
band. Since an exact signed distance field satisfies
$\|\nabla\varphi\|=1$ away from non-differentiable locations, lower residuals
indicate closer agreement with signed-distance behaviour. These quantities
are reported as stratum-level and corpus-level summaries rather than as
per-scene acceptance criteria.

The audit summary is given in Table~\ref{tab:sdf_corpus_fidelity}, while the full per-family breakdown is reported in Appendix Table~\ref{tab:sdf_corpus_fidelity_all}. Across the full subset, the corpus-mean surface distance is $\mathrm{MSD}/\Delta x = 0.283$, and the mean Hausdorff distance is $H/\Delta x = 1.326$. All single-object families remain clearly below the sub-voxel criterion, with $\mathrm{MSD}/\Delta x$ between $0.23$ and $0.29$. Multi-object scenes show larger deviations, especially in the Hausdorff distance, as expected from packed objects and sharp local features. However, the average surface error remains below one voxel even for three-object scenes, where $\mathrm{MSD}/\Delta x = 0.82$.

The aggregate sign and distance-field diagnostics support the same
conclusion. Across the evaluated subset, the corpus-mean sign consistency
is $99.944\%$, and the mean eikonal residual is $0.0305$. The corresponding
stratum-level values also remain similar across the six shape families and
the multi-object configurations. These results indicate that the generated
fields are correctly signed to a high degree and closely approximate
signed-distance behaviour on average across the evaluated subset.

Overall, the production-geometry audit shows sub-voxel average surface
alignment across the evaluated subset. The larger worst-case deviations in
multi-object scenes are local, while the mean reconstructed boundary remains
within one voxel. Together with the aggregate sign-consistency and eikonal
results, this supports the use of the evaluated SDFs as geometric inputs for
the downstream surrogate demonstration.

\begin{table}[pos=t]
\centering
\caption{Corpus-scale SDF fidelity summary for the separate 350-pair SDF verification subset at $256\times128\times128$ ($\Delta x=4$~lu). Single-object values are reported as the range across the six shape families.}
\label{tab:sdf_corpus_fidelity}
\begin{tabular}{l c c c c c}
\toprule
Stratum & $N$ & $\mathrm{MSD}/\Delta x$ & $H/\Delta x$ & Sign consistency (\%) & Eikonal residual \\
\midrule
Single-object families & $300$ & $0.23$--$0.29$ & $0.98$--$1.35$ & $99.884$--$100.000$ & $0.0218$--$0.0427$ \\
Multi-object (2 obj.)  & $36$  & $0.35$ & $1.82$ & $99.923$ & $0.0304$ \\
Multi-object (3 obj.)  & $14$  & $0.82$ & $3.99$ & $99.973$ & $0.0306$ \\
Corpus                 & $350$ & $0.283$ & $1.326$ & $99.944$ & $0.0305$ \\
\bottomrule
\end{tabular}
\end{table}

\subsection{Lattice-Boltzmann Solver Validation}
\label{sec:solver_validation}

The solver is assessed at two complementary levels. First,
Section~\ref{sec:solver_validation_sphere} compares the configured
solver setup with canonical sphere-flow reference data. Second,
Section~\ref{sec:solver_data_integrity} evaluates the final packaged
simulations through a per-scene data-integrity audit across the released
corpus.

\subsubsection{Validation against canonical sphere flow}
\label{sec:solver_validation_sphere}

The configured solver setup is assessed using canonical flow past a centrally placed sphere. Wake statistics are validated against the experimental measurements of Choi and Park~\cite{choi2018flow} at a sphere Reynolds number $Re_s = 1000$, defined as $Re_s = U_\infty D / \nu$ with sphere diameter $D$; this corresponds to a channel-height-based Reynolds number $Re_c \approx 6000$. Drag coefficients are additionally compared with the measurements of Roos and Willmarth~\cite{roos1971some} over the wider range $Re_c = 600$ to $12\,000$. The sphere is placed at one-third of the streamwise domain length in a $2 \times 1 \times 1$ channel with a single-object configuration; all other solver parameters match Section~\ref{sec:solver_lbm} exactly. Reference curves were digitised from the published figures using WebPlotDigitizer v5
~\cite{rohatgi2024webplotdigitizer} where numerical tables were unavailable.

\begin{figure}[pos=htbp]
  \centering
  \includegraphics[width=\columnwidth]{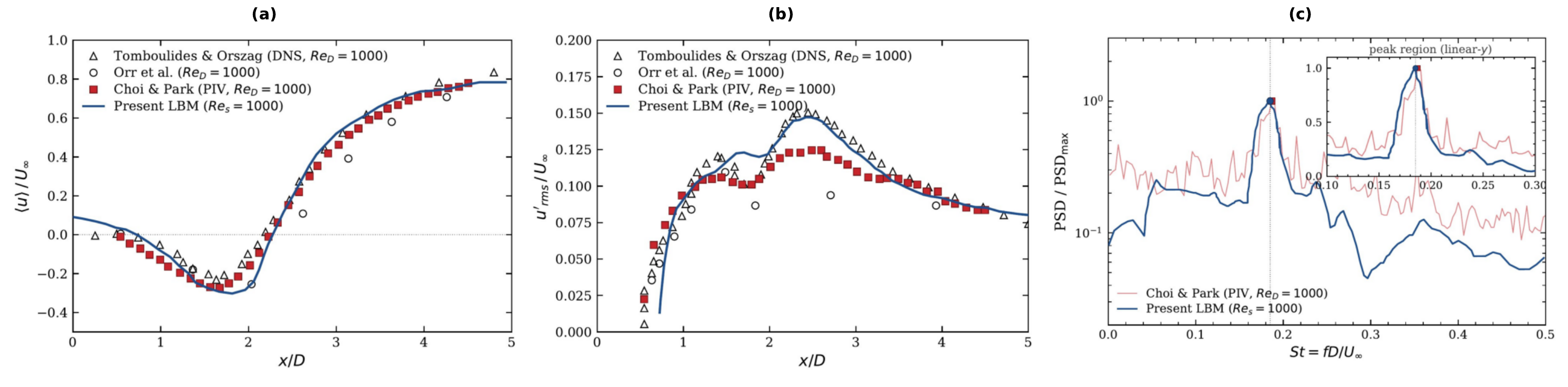}
  \caption{Wake characteristics behind a single sphere at $Re_s = 1000$: 
  (a) normalised mean streamwise velocity along the centreline,
  $\langle u \rangle / U_\infty$; 
  (b) streamwise turbulence intensity along the centreline,
  $u'_{rms} / U_\infty$; 
  (c) power spectral density of the vertical velocity.}
  \label{fig:validation_sphere_re1000}
\end{figure}

Figure~\ref{fig:validation_sphere_re1000} compares the principal mean and
unsteady wake quantities. The computed recirculation length is
$l_r/D\approx1.8$, consistent with the experimental value
$1.76\pm0.07$ reported by Choi and Park~\cite{choi2018flow} and within the
range of the additional numerical references~\cite{tomboulides2000numerical,orr2015numerical}. The streamwise
turbulence-intensity profile shows two main peaks near $x/D\approx1.5$ and
$x/D\approx2.4$. Their locations agree with the reference data, while their
magnitudes remain within the range formed by the PIV and DNS results. The
power spectral density has a dominant peak at
$St=fD/U_\infty\approx0.18$, in agreement with published sphere-wake
measurements and simulations
~\cite{choi2018flow,sakamoto1990study,tomboulides2000numerical}. These
comparisons show that the solver reproduces the principal mean-flow,
fluctuation, and vortex-shedding quantities of the benchmark case.

The drag coefficient provides an integrated assessment of the pressure,
viscous, and wake contributions and is therefore used to examine grid
convergence~\cite{Roache1997}. Table~\ref{tab:drag_convergence} reports
$C_D$ at four Reynolds numbers and four lattice resolutions. At the
verification resolution of $1024\times512\times512$~lu, all four cases
agree with the measurements of Roos and Willmarth
~\cite{roos1971some} within approximately $6\%$. For
$Re_s\leq1000$, the variation across the three finest grids remains below
$2\%$, indicating that the drag coefficient is effectively resolved at
the production grid. The $Re_s=2000$ case, corresponding to
$Re_c=12\,000$, shows a small non-monotone variation across the finer
grids and is therefore interpreted more cautiously.

\begin{table}[pos=htbp]
  \centering
  \caption{Drag coefficient $C_D$ as a function of Reynolds number and
    lattice resolution (full streamwise $\times$ spanwise $\times$
    wall-normal cell counts in lu). Reference values are from Roos and
    Willmarth~\cite{roos1971some}.}
  \label{tab:drag_convergence}
  \footnotesize
  \setlength{\tabcolsep}{4pt}
  \begin{tabular}{ccccccc}
    \toprule
    $Re_s$ & $Re_c$ & $128 \times 64 \times 64$ & $256 \times 128 \times 128$
    & $512 \times 256 \times 256$ & $1024 \times 512 \times 512$ & Reference \\
    \midrule
    $100$  & $600$    & $1.122$ & $1.108$ & $1.106$ & $1.106$ & $1.08$  \\
    $500$  & $3000$   & $0.573$ & $0.562$ & $0.563$ & $0.562$ & $0.547$ \\
    $1000$ & $6000$   & $0.479$ & $0.460$ & $0.466$ & $0.467$ & $0.472$ \\
    $2000$ & $12000$  & $0.434$ & $0.407$ & $0.407$ & $0.404$ & $0.430$ \\
    \bottomrule
  \end{tabular}
\end{table}

The instantaneous velocity field also shows the unsteady, separated wake
expected for this regime~\cite{sakamoto1990study,johnson1999flow}.

Overall, the canonical sphere benchmark verifies the configured solver
setup for a controlled obstructed-flow case. It does not demonstrate
reference accuracy for every geometry in the released corpus.
Section~\ref{sec:solver_data_integrity} therefore complements this benchmark
with a per-scene audit of the final packaged simulations.

\subsubsection{Per-scene data integrity audit}
\label{sec:solver_data_integrity}

The reference validation in Section~\ref{sec:solver_validation_sphere} shows that the solver setup reproduces known sphere-flow benchmark results for one controlled case. However, this does not automatically guarantee that every production simulation is complete, stable, correctly packaged, and physically consistent. Individual runs can still fail in ways that are not visible from the benchmark alone.

We therefore apply the data-integrity audit to all 450 complete simulation scenes in the release. This audit is new to the present manuscript. Each scene links its obstacle geometry to the SDF generated from that geometry and to the corresponding flow fields. The audit therefore checks the final co-registered data product delivered by the pipeline, including the resampling and packaging steps.

Table~\ref{tab:per_scene_audit} summarises the five audit criteria. Four checks act as pass/fail gates for release eligibility: completeness and stability, mass conservation, averaging convergence, and geometry--flow registration. The fifth check, momentum balance, is reported as an independent diagnostic rather than a release gate. All 450 released scenes pass the four critical gates.

\begin{table}[pos=htbp]
\centering
\caption{Per-scene data integrity audit criteria and results.}
\label{tab:per_scene_audit}
\resizebox{\textwidth}{!}{%
\begin{tabular}{p{0.23\linewidth} p{0.34\linewidth} p{0.27\linewidth} p{0.20\linewidth}}
\toprule
Check & What it verifies & Pass criterion & Result \\
\midrule
Completeness and stability
& Checks that all required fields are present, correctly shaped, finite, and free of numerical energy spikes.
& Complete fields, finite values, no energy spike
& $450/450$ passed \\

Mass conservation
& Checks that mass flux remains consistent across streamwise cross-sections after resampling to the ML grid.
& Mass-flux imbalance $\leq 2\%$
& $450/450$ passed; median $0.28\%$, maximum $0.50\%$ \\

Averaging convergence
& Checks that the time-averaged flow field is statistically settled and not still changing.
& Split-half difference remains within the Reynolds-number-dependent gate
& $450/450$ passed \\

Geometry--flow registration
& Checks that each flow field belongs to the stored obstacle geometry.
& Obstacle location and size inferred from geometry and flow must match
& $450/450$ passed \\

Momentum-balance diagnostic
& Compares drag from the force log with drag inferred from the wake momentum deficit.
& Independent diagnostic, not a release gate
& Median disagreement $\approx 11\%$ \\
\bottomrule
\end{tabular}%
}
\end{table}

\begin{figure}[pos=htbp]
  \centering
  \includegraphics[width=\textwidth]{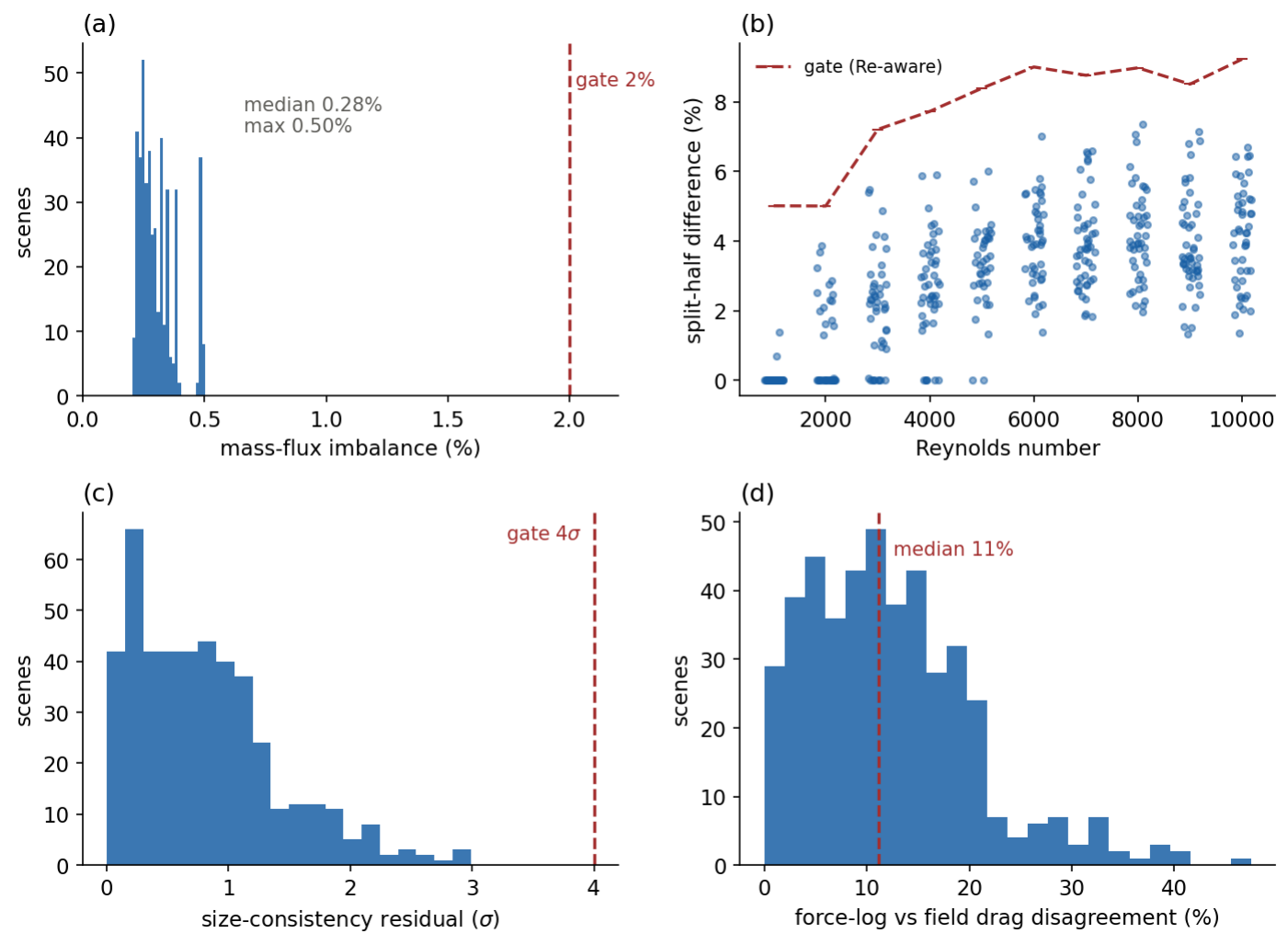}
  \caption{Distribution of the main per-scene audit metrics across the released corpus. 
  (a) Mass-flux imbalance, with the $2\%$ release gate marked by the dashed red line. 
  (b) Split-half averaging difference as a function of Reynolds number, with the Reynolds-number-dependent gate. 
  (c) Geometry--flow size-consistency residual, with the $4\sigma$ gate marked by the dashed red line. 
  (d) Disagreement between force-log drag and wake-implied drag, with the median value marked.}
  \label{fig:per_scene_audit}
\end{figure}

Figure~\ref{fig:per_scene_audit} shows the distribution of the main audit metrics across the corpus. Across the corpus, each gated metric remains well within its threshold. The mass-flux imbalance has a median value of $0.28\%$ and a maximum value of $0.50\%$, well below the $2\%$ gate. The split-half averaging differences remain below the Reynolds-number-dependent gate, showing that the released mean fields are statistically settled. The geometry--flow registration residuals also remain below the acceptance threshold.

The momentum-balance diagnostic provides an additional physics check. It compares the drag recorded in the force log with the drag inferred independently from the wake momentum deficit. These two estimates are computed from different data sources, so their agreement is not circular. The median disagreement is about $11\%$, which is consistent with the accuracy expected from coarse-grid integration. This supports the consistency between the force records and the released flow fields.

Overall, the audit shows that every released scene is complete, numerically stable, mass-conserving, statistically converged, and correctly paired with its geometry. The independent drag diagnostic further supports the physical consistency of the force and field data. The released simulations can therefore be used for downstream surrogate-model training without requiring users to repeat the per-scene verification.

\FloatBarrier

\section{Downstream Surrogate Demonstration}
\label{sec:surrogate_demonstration}

The generated dataset should not only be physically consistent, but also
usable for downstream machine-learning tasks. We therefore train neural
surrogate models to predict the time-averaged flow around an obstacle from
its signed distance field (SDF) and Reynolds number. The aim is not to claim state-of-the-art
surrogate accuracy, but to demonstrate that the generated scenes support
supervised learning and controlled studies of data scaling, shape-family
generalisation, and Reynolds-number extrapolation.

Each sample maps a static obstacle description to the corresponding time-averaged velocity field. The input is a two-channel $128 \times 64 \times 64$ volume. One channel contains the signed distance field (SDF), and the second channel contains the Reynolds number, normalised as $\mathrm{Re}_c/10000$ and broadcast over the full grid. The target is the time-averaged velocity field with three components, $(v_x,v_y,v_z)$, on the same grid. Inputs and targets are normalised using statistics computed only from the training split of each experiment. The experiments use 450 scenes covering six obstacle families, namely cone, cuboid, cylinder, sphere, torus, and wedge, and ten channel-based Reynolds numbers from $Re_c = 1{,}000$ to $10{,}000$. The surrogate experiments use the $128 \times 64 \times 64$ release tier; the same 450 scenes are also released at $256 \times 128 \times 128$.

We compare three surrogate architectures under the same training setup, namely a 3D U-Net ~\cite{ronneberger2015_unet}, a Fourier neural operator (FNO) ~\cite{Li2021FNO}, and a U-Net-enhanced
Fourier neural operator (U-FNO) ~\cite{wen2022_ufno}. The models are trained using mean-squared
error on the normalised velocity components and are evaluated using
relative $L_2$ error after transforming the predictions back to the original
velocity scale. Error is evaluated over the full domain and over a
geometry-dependent region containing the obstacle and its downstream wake.
The wake-region error is used as the primary interpretive metric because
most of the remaining domain contains comparatively smooth channel flow.

Each model configuration is trained once using the fixed random seed 42.
Complete details of the dataset splits, input processing, model architectures,
training configuration, checkpoint selection, evaluation metrics, and
computational environment are provided in the companion data and code
archives. The model architectures, training setup, and computational
resources are summarised in \ref{app:surrogate_reproducibility},
with model architecture provided in
Figs.~\ref{fig:unet3d_architecture},
\ref{fig:fno_architecture}, and
\ref{fig:ufno_architecture}.

\subsection{Corpus-size learning curve}
\label{sec:surrogate_learning_curve}

The first experiment checks how surrogate accuracy changes with training-set size. Each model is trained with 100, 200, and 300 scenes and evaluated on the same fixed validation and test set. The scene identifiers in the training, validation, and test sets are disjoint, and the smaller training subsets are contained within the larger ones. Exact split membership is provided in released manifests. \cite{channelflowtools_software}

\begin{figure}[pos=htbp]
  \centering
  \includegraphics[width=\textwidth]{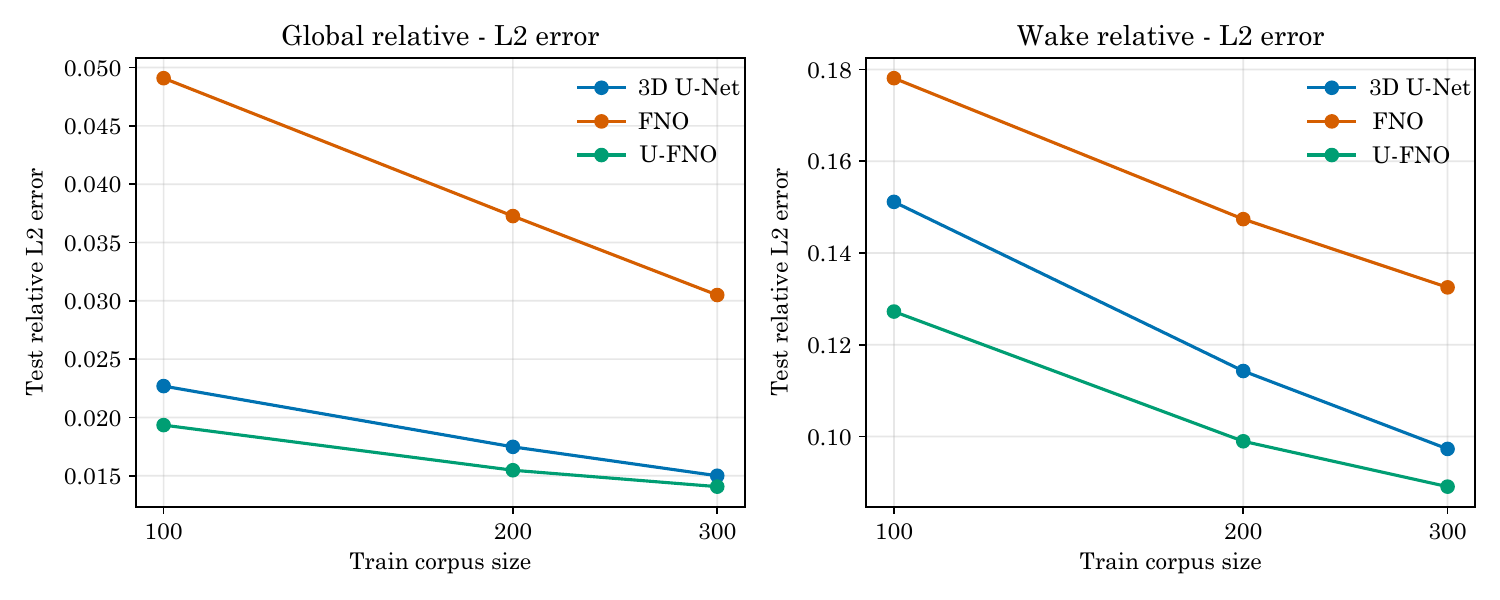}
  \caption{Corpus-size learning curve. Test relative-$L_2$ error versus training corpus size for the three surrogate models. The left panel reports global error over the full domain, while the right panel reports wake-region error.}
  \label{fig:surrogate_learning_curve}
\end{figure}

Figure~\ref{fig:surrogate_learning_curve} shows that all three models improve as more training data are added, in both global and wake-region error. The wake error remains clearly larger than the global error, making it the more informative metric for judging obstacle-induced flow prediction. Since the curves are still decreasing at 300 scenes, the models have not yet saturated, suggesting that additional audited scenes would likely improve performance further.

Across all training sizes, U-FNO gives the lowest wake error, followed closely by U-Net, while FNO remains less accurate. Overall, the experiment shows that the dataset provides a learnable geometry-to-flow mapping and that surrogate performance improves consistently with more data.

\FloatBarrier
\subsection{Shape-family generalisation}
\label{sec:surrogate_shape_ood}

The second experiment tests whether the surrogates can predict obstacle families that were not seen during training. We use a leave-one-family-out setup: each model is trained on five shape families and tested on the remaining held-out family. The held-out wake error is then compared with the model's in-distribution reference.

\begin{figure}[pos=htbp]
  \centering
  \includegraphics[width=\textwidth]{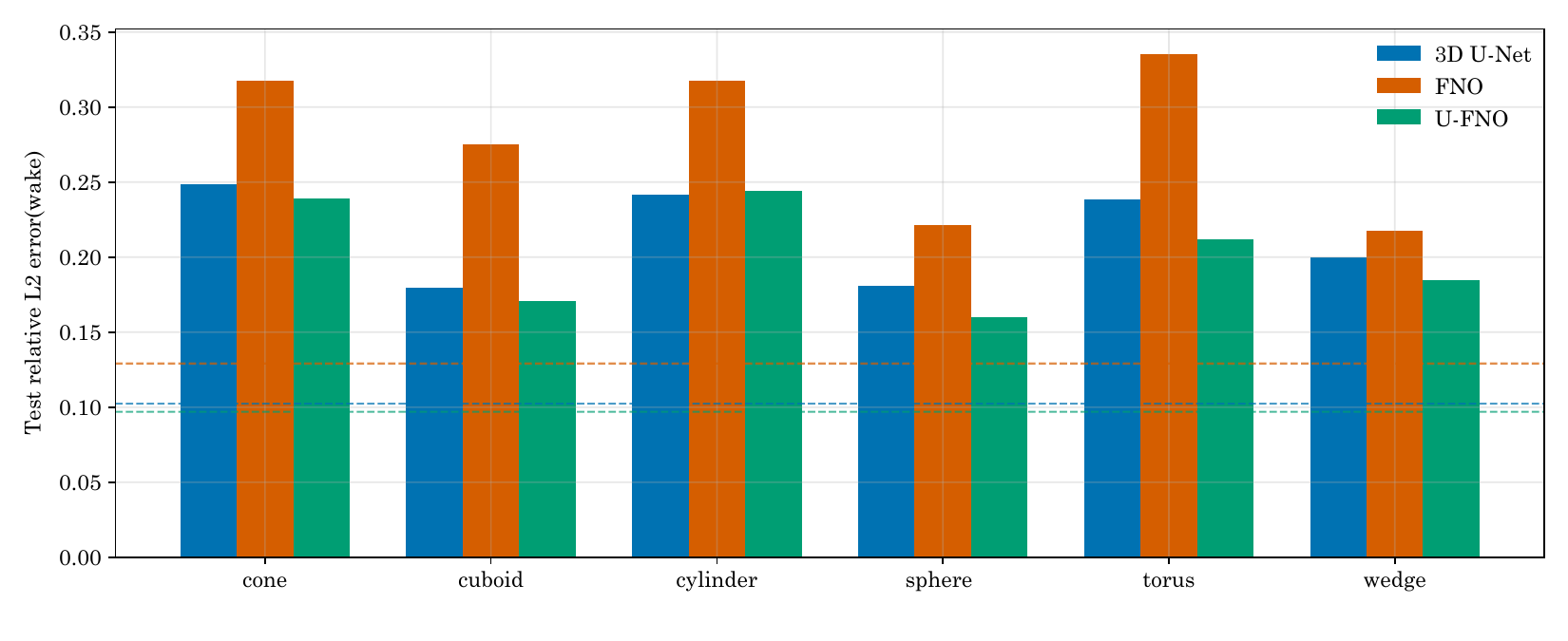}
  \caption{Shape-family out-of-distribution test. Each group corresponds to one held-out obstacle family, and each bar reports the wake relative-$L_2$ error for one model. Dashed horizontal lines show the corresponding in-distribution reference error for each model.}
  \label{fig:surrogate_shape_ood}
\end{figure}

Figure~\ref{fig:surrogate_shape_ood} shows that unseen-shape prediction is
consistently harder than in-distribution prediction. For every model and
every held-out family, the OOD wake error is higher than the corresponding
in-distribution reference. Split separation is established directly from
the scene manifests. For each experiment, the held-out family is absent
from both the training and validation sets and appears only in the test
set.

The hardest case is FNO tested on the held-out torus family, while the easiest case is U-FNO tested on the held-out sphere family, although the sphere result is less stable because it contains only eight test scenes. The model ranking remains consistent with the learning-curve experiment: U-FNO and U-Net perform best, while FNO is less accurate.

\FloatBarrier
\subsection{Reynolds-number extrapolation}
\label{sec:surrogate_re_ood}

The third experiment tests whether the surrogates can extrapolate beyond the Reynolds-number range seen during training. The models are trained only on mid-range cases, $Re_c = 3000$ to $8000$, and then evaluated on lower and higher ranges. The low-Reynolds test set contains cases with $Re_c \leq 2000$, while the high-Reynolds test set contains cases with $Re_c \geq 9000$.

\begin{figure}[pos=htbp]
  \centering
  \includegraphics[width=\textwidth]{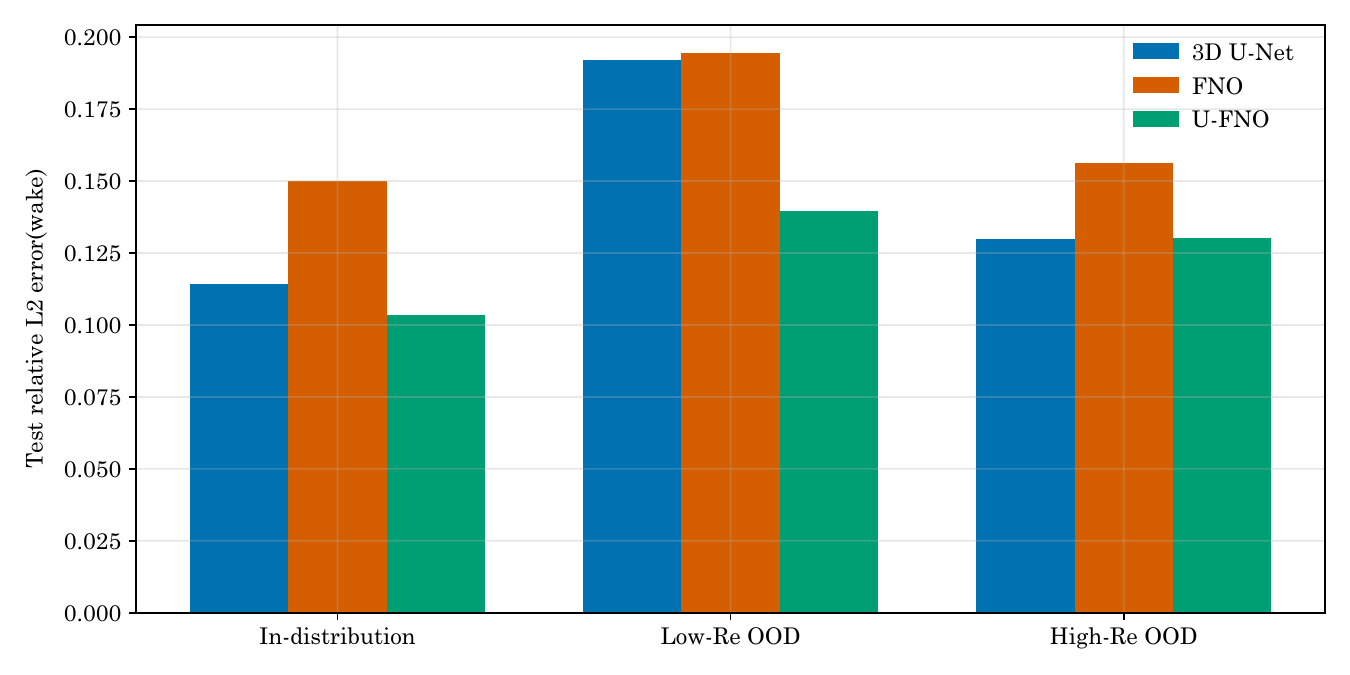}
  \caption{Reynolds-number out-of-distribution test. Bars show wake relative-$L_2$ error for in-distribution validation, low-Reynolds extrapolation, and high-Reynolds extrapolation.}
  \label{fig:surrogate_re_ood}
\end{figure}

Figure~\ref{fig:surrogate_re_ood} shows that both extrapolation ranges are harder than the in-distribution validation case. The low-Reynolds range is the more difficult direction, with wake errors about 1.3 to 1.7 times higher than the in-distribution values. This is physically reasonable because the low-Reynolds cases have flow structures that differ more strongly from the mid-range training data. The high-Reynolds cases also increase the error, but the degradation is smaller.

\section{Conclusion, Limitations, and Future Work}
\label{sec:conclusion_future_work}

\subsection{Conclusion}

We introduced ChannelFlow-Tools, an open and configuration-driven pipeline for generating three-dimensional obstructed-channel-flow datasets. The pipeline covers the full workflow: procedural obstacle generation, signed-distance-field generation, lattice-Boltzmann simulation with waLBerla, and ML-ready tensor packaging on co-registered Cartesian grids.

Each stage of the pipeline was verified independently. The geometry stage was checked through mesh-integrity, reproducibility, scale-stability, and coverage analyses. The SDF stage was validated against analytical primitives and audited against the source STL geometry, showing sub-voxel average surface accuracy at the simulation release tiers. The LBM solver protocol reproduced canonical sphere-flow benchmarks, including wake statistics, vortex shedding, and drag behaviour. The released simulations also passed a per-scene data-integrity audit, confirming that the final ML-ready tensors are complete, stable, mass-conserving, statistically settled, and correctly paired with their geometry.

The primary release contains 450 complete simulation scenes. Each scene is released at both $128 \times 64 \times 64$ ($\Delta x = 8$~lu) and $256 \times 128 \times 128$ ($\Delta x = 4$~lu), and links its obstacle geometry to the corresponding signed distance field and flow fields on co-registered ML grids, providing ready-to-use tensors for downstream learning tasks. The released simulations cover ten channel-based Reynolds numbers from $\mathrm{Re}_c = 1{,}000$ to $10{,}000$ and are used in the surrogate experiments. To support verification and reproduction of the pipeline results, the release also includes a geometry corpus of 10,060 audited scenes across six shape families, with one to three obstacles per scene, and a separate SDF verification subset of 350 geometry--SDF pairs released at $256 \times 128 \times 128$ ($\Delta x = 4$~lu) for the corpus-scale SDF validation. Configuration files, random seeds, metadata, and SHA-256 hash manifests are also provided to support regeneration, verification, and extension of the released resources.

The surrogate experiments show that the released tensors are directly usable for supervised learning. Three baseline models, a 3D U-Net, FNO, and U-FNO, were trained to learn geometry-to-flow mapping. Their errors decrease consistently as the training set grows, showing that the dataset contains a learnable signal. The shape-family and Reynolds-number out-of-distribution experiments show physically interpretable degradation rather than uncontrolled failure. Together, these results show that ChannelFlow-Tools provides a pipeline for studying geometry-aware CFD surrogate models.

\subsection{Limitations}

As with any curated simulation framework, the present release is limited in scope, and each pipeline stage carries its own specific limitations. The geometry generator currently supports six parametric shape families including cone, cuboid, cylinder, sphere, torus, and wedge. Although the generator is designed to be extended, irregular shapes, imported meshes, and non-parametric geometries have not yet been tested.

The mesh-integrity audit also shows local tessellation artefacts near parametric singularities, such as sphere poles and cone apices. These artefacts do not prevent SDF generation or downstream simulation, but they contribute to the STL tessellation error floor observed in the SDF convergence study. Finer tessellation settings may reduce this floor, but this still needs to be tested systematically.

The SDF validation is also bound by the geometries used in this release. Single-object families show clear sub-voxel average surface accuracy, while multi-object scenes show larger deviations, especially for three-object cases. This accuracy is sufficient for the present corpus, but densely packed obstacles and more complex imported geometries may require additional SDF validation.

The solver validation is based on a canonical sphere-flow benchmark and is supported by the per-scene data-integrity audit across the released corpus. This gives confidence in the configured solver protocol, but it does not replace full reference validation for every shape family, obstacle position, or Reynolds-number regime. Additional benchmark cases for other regular obstacles and off-centre placements would further strengthen the validation.

Finally, the surrogate demonstration is intended as a usability study, not as an optimised benchmark of state-of-the-art models. The experiments use a moderate-size corpus, light hyperparameter tuning, and a limited set of baseline architectures. These experiments characterise learning and generalisation difficulty, but they do not claim to provide statistically definitive model comparisons or to solve the generalisation problem.

\subsection{Future Work}

The limitations above suggest several directions for future work. On the generation side, the pipeline can be extended to irregular and non-parametric shape families, such as imported meshes and composite obstacles. Future work will test finer tessellation settings to reduce the STL error floor observed in the SDF validation. Densely packed multi-object scenes and more complex geometries will also require extended SDF evaluation.

On the simulation side, future work will extend validation for more obstacle types and for obstacles placed at different positions inside the channel. Extending the dataset to higher Reynolds numbers will also require finer grid resolutions and renewed solver validation. A systematic study of resampling errors between the lattice grid and the ML grid would further strengthen the reliability of the released tensors. 

On the learning side, the unsaturated learning curves suggest that larger datasets would continue to improve surrogate performance. Future work will therefore scale the released simulation set and define broader standardised benchmark splits for geometry-family generalisation, Reynolds-number extrapolation, and resolution-sensitivity studies. Since the release includes verification tools, configurations, and reproducibility checks, third parties can extend the dataset while preserving the same audit and reproducibility guarantees.

Overall, ChannelFlow-Tools is intended to serve as shared and auditable infrastructure for geometry-aware CFD surrogate research.

\section*{Declaration of Competing Interest}
The authors declare that they have no known competing financial interests or
personal relationships that could have appeared to influence the work reported
in this paper.

\section*{Code and Data Availability}

The ChannelFlow-Tools source code is archived on Zenodo
(\href{https://doi.org/10.5281/zenodo.21806090}{https://doi.org/10.5281/zenodo.21806090})
\cite{channelflowtools_software}.
The generated dataset of 3D obstructed channel-flow simulations is archived on Zenodo
(\href{https://doi.org/10.5281/zenodo.21794316}{https://doi.org/10.5281/zenodo.21794316})
\cite{channelflowtools_dataset}. Both records are currently under restricted access
and will be made openly available upon journal acceptance. Access for research
purposes may be requested from the corresponding author.

\section*{Funding}

\noindent\textbf{Funded by the European Union.} This work has received funding
from the European High Performance Computing Joint Undertaking (JU) and Poland,
Germany, Spain, Hungary, France, Greece under grant agreement number: 101093457.

\noindent This publication expresses the opinions of the authors and not
necessarily those of the EuroHPC JU and Associated Countries which are not
responsible for any use of the information contained in this publication.

\begin{center}
\eulogo{EU_1}\hspace{1.5em}\eulogo{EU_2}
\end{center}

\noindent\textbf{Disclaimer}\\
Funded by the European Union. Views and opinions expressed are however those of
the author(s) only and do not necessarily reflect those of the European Union or
the European High Performance Computing Joint Undertaking (JU) and Poland,
Germany, Spain, Hungary, France, Greece. Neither the European Union nor the
granting authority can be held responsible for them.

\printcredits

\appendix
\numberwithin{table}{section}
\numberwithin{figure}{section}
\renewcommand{\thesection}{Appendix~\Alph{section}}
\renewcommand{\thetable}{\Alph{section}.\arabic{table}}
\renewcommand{\thefigure}{\Alph{section}.\arabic{figure}}
\newpage

\section{Geometry Generation: Dataset Card and Reproducibility Parameters}
\label{app:geometry}

\begin{table}[H]
\centering

\caption{Dataset card for the released \textsc{ChannelFlow-Tools} resources.
The release comprises a geometry corpus, an SDF verification subset, and a
complete simulation corpus. The geometry stage of any released scene can be
regenerated byte-identically from its configuration, seed, and scene index and
verified against the published SHA-256 manifests.}

\label{tab:dataset_card}

\small
\begin{tabularx}{\columnwidth}{@{}l>{\raggedright\arraybackslash}X@{}}
\toprule
\textbf{Field} & \textbf{Value} \\
\midrule

Generated geometry corpus
& 10,060 scenes (15,562 objects); STL meshes and YAML metadata \\

SDF validation subset
& 350 geometry--SDF pairs used specifically for SDF validation
at $256\times128\times128$ ($\Delta x=4$ lu) \\

Complete simulation corpus
& 450 scenes released at both $128\times64\times64$ ($\Delta x=8$ lu) and
$256\times128\times128$ ($\Delta x=4$ lu); each scene contains the geometry
representation, corresponding SDF, and corresponding flow simulation \\

Shape families
& 6: sphere, cuboid, cylinder, cone, torus, wedge \\

Objects per scene
& 1--3; all 83 unordered shape compositions realised \\

Sampling
& Stratified Sobol; seeded uniform sampling also supported \\

Reynolds-number range
& $Re_c = 1000$--$10{,}000$ for the released simulation corpus
(channel-height based) \\

Domain size
& $1024\times512\times512$ lu
(streamwise $\times$ wall-normal $\times$ spanwise) \\

ML grid resolution
& $128\times64\times64$ ($\Delta x=8$ lu) and $256\times128\times128$ ($\Delta x=4$ lu);
other resolutions regenerable from configuration \\

Stored variables
& Geometry representation, signed SDF $\phi$, and time-averaged velocity
components $(v_x,v_y,v_z)$; obstacle force logs \\

File formats
& HDF5 for complete simulation scenes; STL and YAML for generated geometries \\

Metadata fields
& Shape family, sampled parameters, placement, orientation, seed, scene index,
and mesh-quality statistics \\

Reproducibility
& Configuration + seed + scene index; SHA-256 hash manifests;
byte-identical geometry regeneration
(\S\ref{sec:v6_reproducibility}) \\

Train/validation/test splits
& Configurable; corpus-size study uses a 325-scene training pool,
46-scene fixed validation set, and 79-scene fixed test set, with nested
training subsets of 100, 200, and 300 scenes; shape-family-OOD and
Reynolds-OOD splits also provided \\

Licence
& The ChannelFlow-Tools software is released under the GPL-3.0 licence, while the accompanying dataset is released under CC BY 4.0 licence.
 \\

Code (Zenodo)
& \href{https://doi.org/10.5281/zenodo.21806090}
{10.5281/zenodo.21806090} \\

Dataset DOI
& \href{https://doi.org/10.5281/zenodo.21794316}
{10.5281/zenodo.21794316} \\

\bottomrule
\end{tabularx}
\end{table}

\begin{table}[pos=H]
\centering
\footnotesize
\setlength{\tabcolsep}{6pt}
\renewcommand{\arraystretch}{1.2}
\caption{Key configuration parameters for reproducing the ChannelFlow-Tools
pipeline. Values not listed here are fixed in the released configuration files.}
\label{tab:repro-params}
\begin{tabular}{@{}p{0.40\textwidth}p{0.52\textwidth}@{}}
\toprule
\textbf{Parameter} & \textbf{Value / setting} \\
\midrule
\multicolumn{2}{@{}l}{\textit{Geometry generation}} \\
\midrule
Shape families              & sphere, cuboid, cylinder, cone, torus, wedge \\
Objects per scene           & 1--3 \\
Canonical domain            & $1024\times512\times512$~lu (centre $[512,256,256]$) \\
Sampler                     &  Sobol (scipy \texttt{qmc.Sobol}) \\
Inlet/outlet exclusion      & $L_u = L_d = 100$~lu \\
Wall clearance              & 20~lu \\
STL tessellation (lin./ang.)& $1.0\times10^{-4}$ (relative) / $0.05$~rad \\
Mesh format                 & STL, one scene-level file per scene \\
\midrule
\multicolumn{2}{@{}l}{\textit{SDF generation}} \\
\midrule
Library                     & OpenVDB (\texttt{createLevelSetFromPolygons}), $\geq 12.1$ \\
Sign convention             & signed distance $\varphi$; $\varphi<0$ inside, $\varphi>0$ outside \\
Simulation release tiers    & $\Delta x = 8$~lu $\rightarrow 128\times64\times64$;
$\Delta x = 4$~lu $\rightarrow 256\times128\times128$ \\
SDF verification subset     & 350 geometry--SDF pairs at
$256\times128\times128$ ($\Delta x = 4$~lu) \\
Validation ladder           &
$64\times32\times32$, 
$128\times64\times64$, $256\times128\times128$, $512\times256\times256$,
$1024\times512\times512$ \\
\midrule
\multicolumn{2}{@{}l}{\textit{Lattice-Boltzmann simulation (waLBerla, block-structured)}} \\
\midrule
Velocity stencil            & D3Q27 \\
Collision operator          & cumulant (compressible, zero-centred) \\
Lattice velocity $U_\mathrm{LB}$ & $0.05$ \\
Viscosity / relaxation      & $\nu = U_\mathrm{LB} L_\mathrm{ref}/Re$; $\omega = 1/(3\nu+0.5)$ \\
Reynolds number &
Channel-height based, $Re_c=U_\infty H/\nu$;
configured range $Re_c\approx100$--$15{,}000$,
released corpus $Re_c=1{,}000$--$10{,}000$,
and sphere benchmark $Re_c=600$--$12{,}000$ \\
Inlet boundary              & uniform velocity (UBB) \\
Outlet boundary             & extrapolation outflow \\
Wall boundary               & no-slip (linear bounce-back) \\
Obstacle boundary           & quadratic bounce-back \\
Production domain           & $1024\times512\times512$~lu (streamwise $\times$ wall-normal $\times$ spanwise) \\
\midrule
\multicolumn{2}{@{}l}{\textit{ML tensor packaging}} \\
\midrule
Resampling                  & trilinear (scipy \texttt{RegularGridInterpolator}), $8\times$ downsampling ($\Delta x = 8$~lu tier) and $4\times$ downsampling ($\Delta x = 4$~lu tier) \\
Target grids                & $128\times64\times64$ ($\Delta x = 8$~lu) and
$256\times128\times128$ ($\Delta x = 4$~lu) \\
Stored fields &
geometry field; signed SDF $\varphi$; time-averaged velocity components
$(v_x,v_y,v_z)$; optional instantaneous velocity and density
snapshots \\
Auxiliary output &
obstacle force log \\
File format                 & HDF5, one file per scene \\
\bottomrule
\end{tabular}
\end{table}

\FloatBarrier
\section{Signed Distance Field Generation: Detailed Reconstruction Results}
\label{app:sdf}
\begin{figure}[pos=H]
\centering
\includegraphics[width=\textwidth]{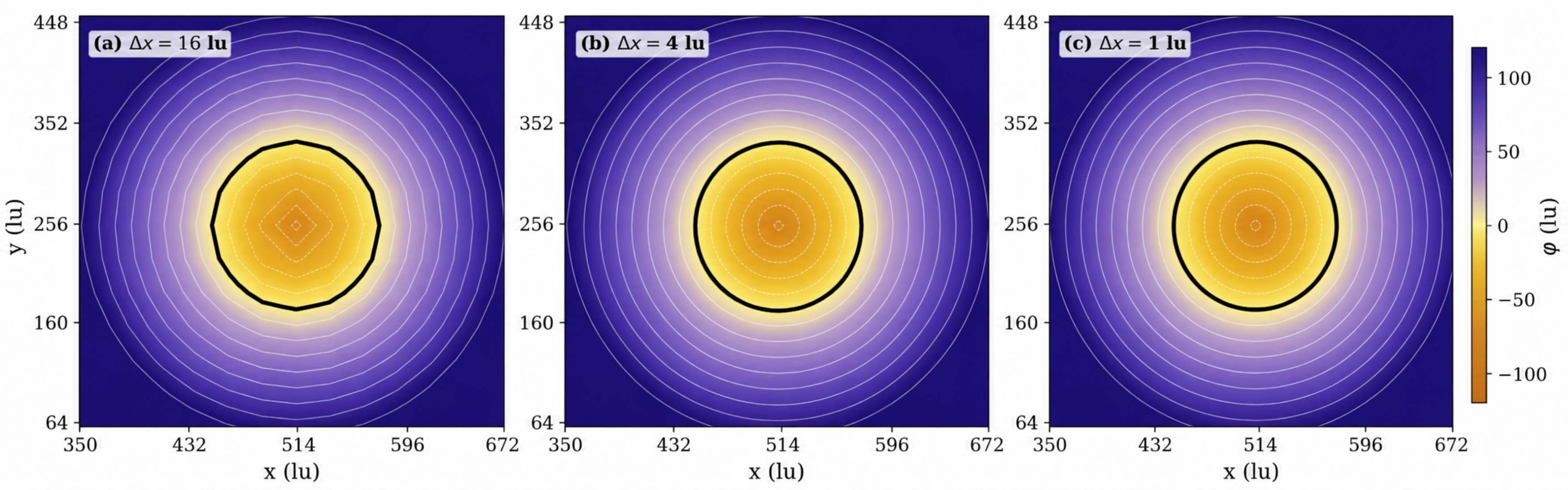}
\caption{Mid-plane cross-section of the sphere SDF ($r = 80$~lu) at three
  grid resolutions: (a)~$64 \times 32 \times 32$ ($\Delta x = 16$~lu),
  (b)~$256 \times 128 \times 128$ ($\Delta x = 4$~lu), and
  (c)~$1024 \times 512 \times 512$ ($\Delta x = 1$~lu), zoomed to the
  obstacle region. The zero-level-set (thick black contour) transitions from
  a visibly faceted polygon at coarse resolution to a smooth circle at fine
  resolution.}
\label{fig:sphere_multiresolution}
\end{figure}

\begin{table}[pos=H]
\centering
\caption{Surface-reconstruction metrics for sphere, cuboid, and cylinder across the five-resolution ladder, mean values over seven realisations per shape family. Chamfer is reported in lu$^{2}$, while Hausdorff distance $H$ and MSD are reported in lu. MSD$/\Delta x$ gives the mean surface distance normalised by the voxel spacing.}
\label{tab:sdf_surface_recon_appendix}
\resizebox{\textwidth}{!}{%
\begin{tabular}{l c c c c c c c c c c c c c}
\toprule
 & & \multicolumn{4}{c}{Sphere} & \multicolumn{4}{c}{Cuboid} & \multicolumn{4}{c}{Cylinder} \\
\cmidrule(lr){3-6} \cmidrule(lr){7-10} \cmidrule(lr){11-14}
Resolution & $\Delta x$ (lu)
& Chamfer & $H$ & MSD & MSD$/\Delta x$
& Chamfer & $H$ & MSD & MSD$/\Delta x$
& Chamfer & $H$ & MSD & MSD$/\Delta x$ \\
\midrule
$64 \times 32 \times 32$
& $16$
& $9.44$ & $5.75$ & $1.94$ & $0.121$
& $15.36$ & $9.49$ & $2.08$ & $0.130$
& $12.66$ & $7.70$ & $2.20$ & $0.138$ \\

$128 \times 64 \times 64$
& $8$
& $7.84$ & $5.70$ & $1.63$ & $0.204$
& $5.93$ & $5.68$ & $1.46$ & $0.183$
& $11.29$ & $7.32$ & $2.03$ & $0.254$ \\

$256 \times 128 \times 128$
& $4$
& $7.75$ & $5.60$ & $1.60$ & $0.400$
& $5.00$ & $4.92$ & $1.35$ & $0.338$
& $11.04$ & $7.06$ & $2.00$ & $0.500$ \\

$512 \times 256 \times 256$
& $2$
& $7.71$ & $5.49$ & $1.59$ & $0.795$
& $4.98$ & $4.92$ & $1.34$ & $0.670$
& $10.93$ & $7.04$ & $1.99$ & $0.995$ \\

$1024 \times 512 \times 512$
& $1$
& $7.70$ & $5.48$ & $1.59$ & $1.590$
& $4.95$ & $4.63$ & $1.34$ & $1.340$
& $10.90$ & $7.01$ & $1.99$ & $1.990$ \\
\bottomrule
\end{tabular}%
}
\end{table}

\begin{table}[pos=H]
\centering
\caption{Corpus-scale SDF fidelity at the verification tier
($\Delta x = 4$~lu), per stratum and corpus-wide. MSD/$\Delta x$ and
Hausdorff/$\Delta x$ are the surface-reconstruction metrics of
Section~\ref{sec:sdf_surface_recon} computed against each scene's STL; sign
consistency is the percentage of narrow-band voxels whose field sign matches
an independent point-in-mesh test; the eikonal residual is the band-mean of
$\lvert \lVert \nabla \varphi \rVert - 1 \rvert$.}
\label{tab:sdf_corpus_fidelity_all}
\begin{tabular}{l c c c c c}
\toprule
Stratum & $N$ & $\mathrm{MSD}/\Delta x$ & $H/\Delta x$ & Sign consistency (\%) & Eikonal residual (mean) \\
\midrule
Cuboid                & $50$  & $0.27$ & $0.98$ & $100.000$ & $0.0302$ \\
Sphere                & $50$  & $0.23$ & $1.04$ & $99.911$  & $0.0261$ \\
Torus                 & $50$  & $0.29$ & $1.10$ & $99.884$  & $0.0218$ \\
Cylinder              & $50$  & $0.25$ & $1.15$ & $99.952$  & $0.0271$ \\
Wedge                 & $50$  & $0.23$ & $1.16$ & $100.000$ & $0.0427$ \\
Cone                  & $50$  & $0.25$ & $1.35$ & $99.927$  & $0.0349$ \\
Multi-object (2 obj.) & $36$  & $0.35$ & $1.82$ & $99.923$  & $0.0304$ \\
Multi-object (3 obj.) & $14$  & $0.82$ & $3.99$ & $99.973$  & $0.0306$ \\
Corpus (350)          & $350$ & $0.283$ & $1.326$ & $99.944$  & $0.0305$ \\
\bottomrule
\end{tabular}
\end{table}

\FloatBarrier
\section{Lattice-Boltzmann Solver: Per-Scene Audit Criteria}
\label{app:solver}

\begin{table*}[t]
\centering
\caption{Per-scene data-integrity audit applied to all 450 released simulations.
$\mathrm{KE}(t)=\mathrm{mean}_{\mathrm{cells}}|\mathbf{u}|^2$ over the 20 stored
frames. The first four criteria are release gates; momentum balance is an
additional consistency diagnostic.}
\label{tab:audit-criteria}

\small

\begin{tabularx}{\textwidth}{
@{}
>{\raggedright\arraybackslash}p{2.15cm}
>{\raggedright\arraybackslash}X
>{\raggedright\arraybackslash}X
@{}
}
\toprule
\textbf{Criterion} &
\textbf{Definition} &
\textbf{Pass criterion / calibration} \\
\midrule

Completeness and stability &
Required fields have the expected shapes and finite values.
A frame is an energy spike if
$|\mathrm{KE}-\mathrm{med}(\mathrm{KE})|>5\,\mathrm{IQR}$ and
$>0.01\,\mathrm{med}(\mathrm{KE})$. &
All structural checks pass and no energy spike occurs. \\

Mass conservation &
Plane-mean streamwise mass flux
$F(x)=\langle \rho u_x\rangle$;
$\varepsilon_F=(\max F-\min F)/|\mathrm{mean}(F)|$. &
$\varepsilon_F\leq2\%$. \\

Averaging convergence &
The 20 frames are split into two halves:
$\varepsilon_A=
|\bar{\mathbf{v}}_{0:10}-\bar{\mathbf{v}}_{10:20}|_2/
|\bar{\mathbf{v}}|_2$. &
$\varepsilon_A\leq\mathrm{baseline}(Re_c)+0.05$.
 \\

Geometry--flow registration &
Obstacle location and size inferred from geometry and flow are compared.
For size,
$\delta=\ln S_f-(1.257\ln S_g-1.037)$,
where $S_g$ and $S_f$ are geometry- and flow-derived size measures. &
Lateral displacement $\leq4$ voxels and
$|\delta|\leq4\sigma$, with
$\sigma=0.444$
($4\sigma=1.776$). \\

Momentum balance (diagnostic) &
Momentum flux
$M(x)=\sum_{y,z}(\rho u_x^2+\rho/3)$ gives the wake-based drag
$D_{\mathrm{field}}$, compared with the force-log value $F_x$ using
$\varepsilon_D=
|F_x-KD_{\mathrm{field}}|/|F_x|$. &
No release gate.
$K=\mathrm{median}(F_x/D_{\mathrm{field}})=51.369$
for the $128\times64\times64$ output grid. \\

\bottomrule
\end{tabularx}
\end{table*}

\FloatBarrier
\section{Downstream Surrogate Demonstration: Splits and Training Configuration}
\label{app:surrogate_reproducibility}

All dataset splits are defined through CSV manifests released with the
experiment code. Each manifest records the scene identifier, obstacle family, geometry variant, Reynolds number, and assigned split. The training, validation, and test scene identifiers were checked for direct overlap before training and were released with manifests \cite{channelflowtools_software}

\begin{table}[pos=H]
\centering
\small
\caption{Model and training configurations used in the surrogate
demonstration.}
\label{tab:surrogate_training}
\begin{tabular}{lrrrrl}
\toprule
Model & Trainable parameters & Batch size & Initial learning rate &
Fourier modes & Main architecture settings \\
\midrule
3D U-Net
& 8,011,907
& 4
& $3\times10^{-4}$
& --
& Feature widths 32, 64, 128, 256 \\

FNO
& 6,561,319
& 4
& $2\times10^{-3}$
& $8\times8\times8$
& Four spectral layers, width 20 \\

U-FNO
& 41,911,827
& 2
& $3\times10^{-4}$
& $8\times8\times8$
& Four Fourier blocks, width 16 \\
\bottomrule
\end{tabular}
\end{table}

All configurations were trained for 200 epochs using Adam with
$\beta_1=0.9$, $\beta_2=0.999$, weight decay $10^{-5}$, cosine learning-rate
decay, and gradient-norm clipping at 1.0. The loss was the mean-squared
error of the normalised velocity components over the complete spatial
domain. Mixed precision was enabled for the 3D U-Net and disabled for the
Fourier models. No data augmentation was applied. Each configuration was trained once using random seed 42. The checkpoint with the lowest validation global relative-$L_2$ error was used for final
evaluation. 
Training was performed on a single NVIDIA A100 GPU with 40~GB memory.
The environment used Python 3.11.14, PyTorch 2.5.1 with CUDA 12.1,
h5py 3.15.1, and NumPy 1.26.4.

\begin{figure}[pos=H]
\centering
\includegraphics[width=\textwidth]{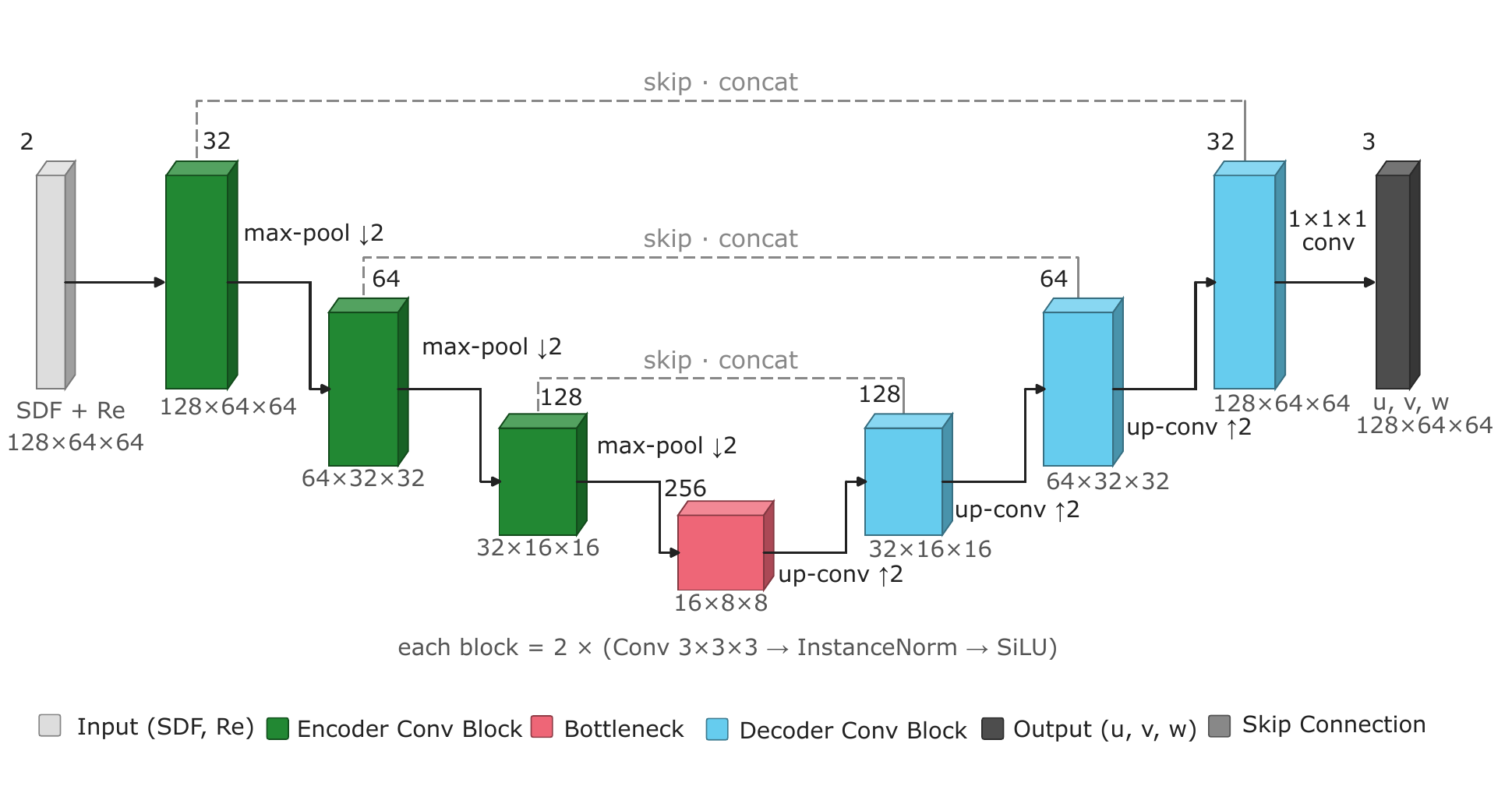}
\caption{3D U-Net model with 3 encoder blocks, a bottleneck, and 3 decoder blocks (feature widths $32/64/128/256$).}
\label{fig:unet3d_architecture}
\end{figure}

\begin{figure}[pos=H]
\centering
\includegraphics[width=\textwidth]{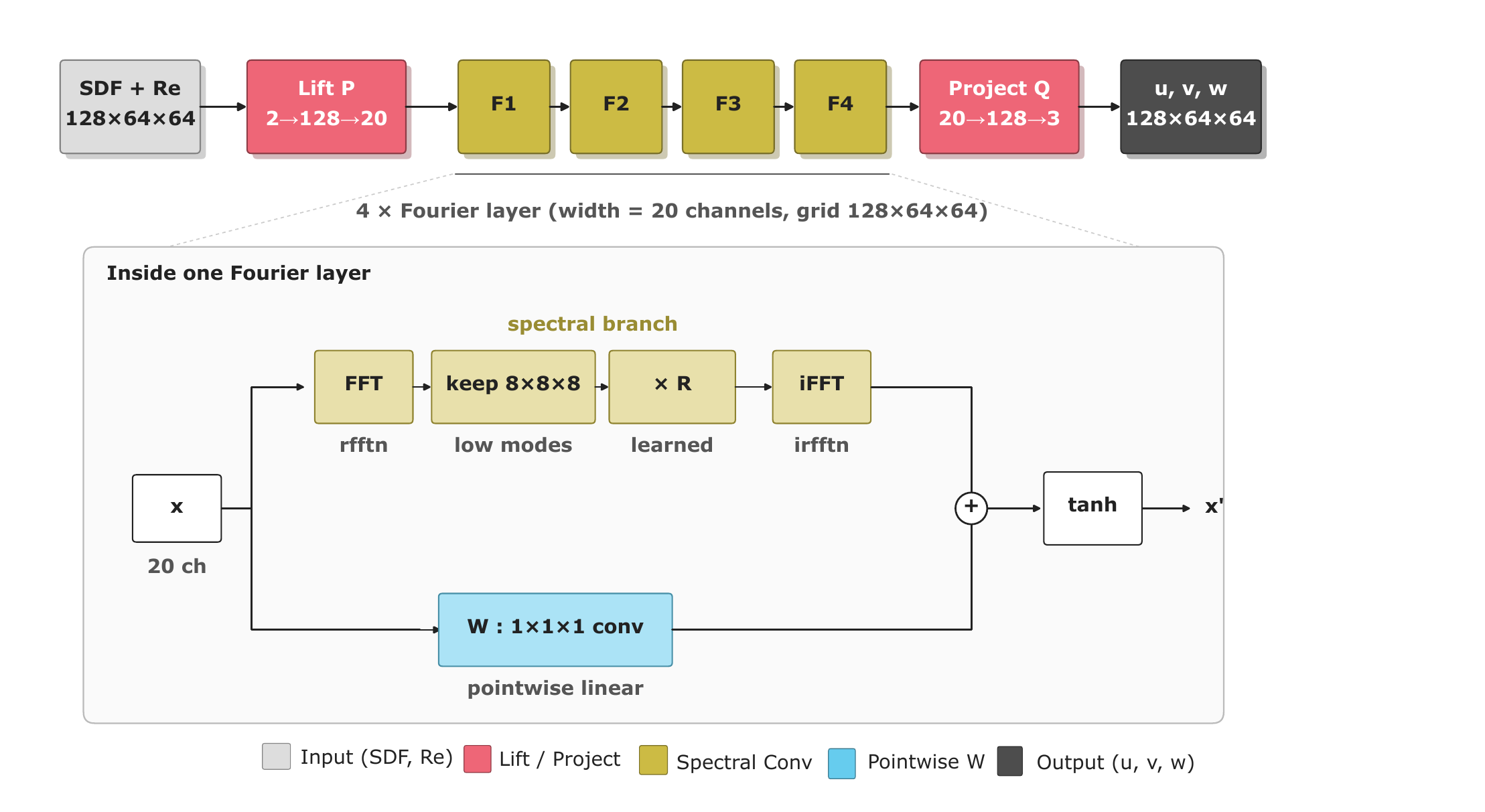}
\caption{FNO model with 4 Fourier blocks and $8\times8\times8$ retained modes (width 20).}
\label{fig:fno_architecture}
\end{figure}

\begin{figure}[pos=H]
\centering
\includegraphics[width=\textwidth]{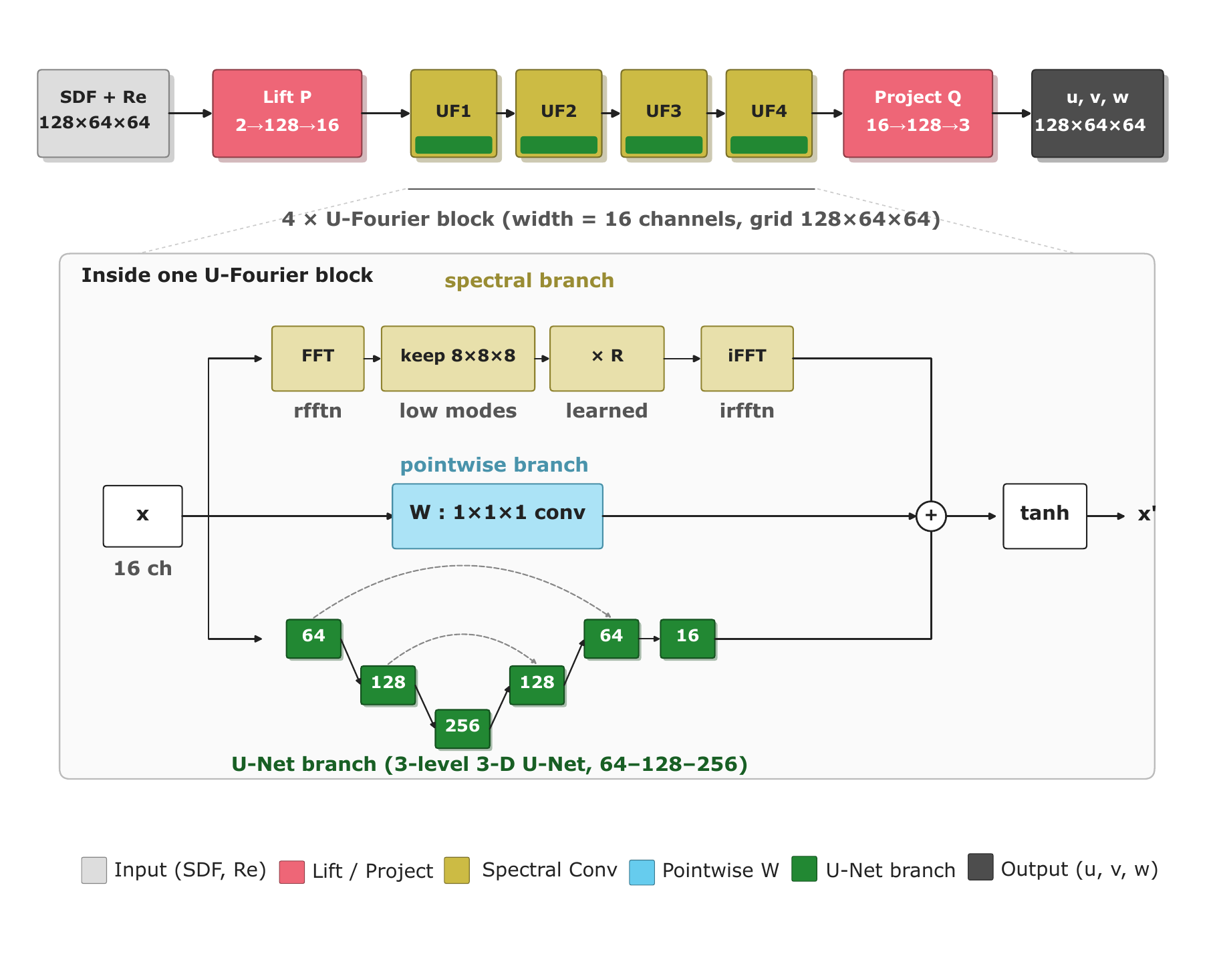}
\caption{U-FNO model with 4 Fourier blocks, each augmented by a U-Net with 3 encoder and 2 decoder stages (width 16).}
\label{fig:ufno_architecture}
\end{figure}

\FloatBarrier
\clearpage

\section*{Acknowledgements}
The authors gratefully acknowledge the scientific support and HPC resources
provided by the Erlangen National High Performance Computing Center (NHR@FAU)
of the Friedrich-Alexander-Universit\"at Erlangen-N\"urnberg. NHR funding is
provided by federal and Bavarian state authorities. NHR@FAU hardware is
partially funded by the German Research Foundation (DFG) -- 440719683.

\bibliographystyle{model1-num-names}
\bibliography{references}

\end{document}